\documentclass[pra,10pt,twocolumn,superscriptaddress]{revtex4-1}
\usepackage{amsmath}
\usepackage{latexsym}
\usepackage{amssymb}
\usepackage{graphicx}
\usepackage{hyperref}
\hypersetup{
    colorlinks = true,
    linkcolor =blue,
	citecolor=blue, 
	urlcolor=blue 
}
\usepackage{mathtools}
\DeclarePairedDelimiter{\ceil}{\lceil}{\rceil}
\begin{document}
\title{Anomalous transport through algebraically localized states in one-dimension}
\author{Madhumita Saha}
\email{madhumita.saha91@gmail.com }
\affiliation{Physics and Applied Mathematics Unit, Indian Statistical
Institute, 203 Barrackpore Trunk Road, Kolkata-700 108, India}
\author{Santanu K. Maiti}
\email{santanu.maiti@isical.ac.in}
\affiliation{Physics and Applied Mathematics Unit, Indian Statistical
Institute, 203 Barrackpore Trunk Road, Kolkata-700 108, India}
\author{Archak Purkayastha}
\email{archakp2@gmail.com}
\affiliation{School of Physics, Trinity College Dublin, College Green, Dublin 2, Ireland}
\begin{abstract}
Localization in one-dimensional disordered or quasiperiodic non-interacting systems in presence of power-law hopping is very different from localization in short-ranged systems. Power-law hopping leads to algebraic localization as opposed to exponential localization in short-ranged systems. Exponential localization is synonymous with insulating behavior in the thermodynamic limit. Here we show that the same is not true for algebraic localization.  We show, on general grounds, that depending on the strength of the algebraic decay, the algebraically localized states can be actually either conducting or insulating in thermodynamic limit. We exemplify this statement with explicit calculations on the Aubry-Andr{\'e}-Harper model in presence of power-law hopping, with the power-law exponent $\alpha>1$, so that the thermodynamic limit is well-defined. We find a phase of this system where there is a mobility edge separating completely delocalized and algebraically localized states, with the algebraically localized states showing signatures of super-diffusive transport. Thus, in this phase, the mobility edge separates two kinds of conducting states, ballistic and super-diffusive. We trace the occurrence of this behavior to near-resonance conditions of the on-site energies that occur due to the quasi-periodic nature of the potential.    
\end{abstract}

\maketitle
\subsection{Introduction}
In the context of disordered non-interacting (quadratic Hamiltonian) systems on a lattice, localization of a single-particle eigenstate refers to the condition where the corresponding eigenfunction has a single highly pronounced peak at a particular system site. The most well-studied form of localization is the Anderson localization \cite{Andersion_random,disordered_electronic}. In one-dimensional short-ranged non-interacting systems, it occurs in presence of a potential with infinitesinmal random disorder. Any single-particle eigenstate of such a system has a pronounced peak at a lattice site, with exponentially decaying tails. This exponential decay allows for definition of a finite single-particle localization length. One of the most important physical effects of such exponential localization is complete absence of transport. In other words, an exponentially localized state is completely insulating in the thermodynamic limit. In contrast, in absence of disorder, all single-particle eigenstates are completely delocalized, leading to ballistic transport.     

Replacing the random disordered potential by a quasi-periodic potential, such as Aubry-Andr{\'e}-Harper (AAH) potential leads to richer physics\cite{Aubry,Harper}. The paradigmatic AAH model consists of a one-dimensional chain with nearest neighbour hopping and the AAH on-site potential. As the strength of the on-site potential is increased, the AAH model shows a phase transition from an all states completely delocalized phase to an all states exponentially localized phase, via a critical point\cite{Aubry}. At the critical point, all states are neither delocalized nor localized but are `critical' or multifractal \cite{Pandit83}. Though the AAH model does not have a mobility edge, slight extensions of the AAH model, such as adding a next nearest neighbour hopping, leads to having mobility edges in energy, separating regions of delocalized and exponentially localized states \cite{Rossignolo,theory_power_law_loc6,ladder1}. The physical effect of having such a mobility edge is that the same system can be conducting or insulating depending on energy. Quasi-periodic systems, with and without mobility edges  are the limelight of recent research \cite{Archak_phase_diagram,Archak_AAH,Archak_AAH1,quasi_periodic_manybody1,
quasi_periodic_manybody,GAAH_mobility_edge,Commensurate_AAH,Correlated_disorder}. These systems have been experimentally realized in several set-ups, with tunable interactions \cite{mobility_edge_expt,I_bloch_experiment,expt1,expt2,expt3,zilberberg1,zilberberg2}. They have got the spotlight recently, with the possibility of exploring the effects of interactions on a system with mobility-edge as one of the main focuses \cite{mobility_edge_expt,machine_learning_mobility_edge, quasi_periodic_manybody1,
quasi_periodic_manybody,GAAH_mobility_edge}. 

Apart from such quasi-periodic systems, a different class of non-interacting systems also show delocalization-localization transitions, as well as possible mobility edges, in 
one-dimension. They are disordered systems with long-ranged hopping which decays as a power-law \cite{theory_old_localization1,theory_old_localization2,theory_old_localization3,
theory_old_localization4,theory_old_localization5,theory_old_localization6,
theory_old_localization7,theory_power_law_loc1,theory_power_law_loc2,theory_power_law_loc3,
theory_power_law_loc4,theory_power_law_loc5,theory_power_law_loc6,
theory_power_law_loc8,theory_power_law_loc9,theory_power_law_loc10,
theory_power_law_loc12,theory_power_law_loc13}. Depending on the power-law decay exponent, the single particle eigenstates of such systems can be delocalized or localized or multifractal. Long range disordered systems have been of growing interest recently due to theoretical and experimental demonstrations of exotic physics in them such as time-crystals ~\cite{expt_time_crystal1,expt_time_crystal2}, prethermalization ~\cite{expt_trapped_ions3,theory_pretherm1,theory_pretherm2,
theory_pretherm3}, dynamical phase transitions~\cite{expt_trapped_ions2,expt_trapped_ions7,expt_atoms_in_trap,theory_DQPT1,theory_DQPT2}, environment assisted transport \cite{Experiment_transport} etc. Recent theoretical exploration into localization properties of long-range systems \cite{theory_old_localization1,theory_old_localization2,
theory_old_localization3,theory_old_localization4,theory_old_localization5,
theory_old_localization6,theory_power_law_loc1,theory_power_law_loc2,theory_power_law_loc3,
theory_power_law_loc4,theory_power_law_loc5,theory_power_law_loc6,
theory_power_law_loc8,theory_power_law_loc9,theory_power_law_loc10,
theory_power_law_loc12,theory_power_law_loc13,
theory_mbl_long_range1,theory_mbl_long_range2,
theory_mbl_long_range3,theory_mbl_long_range4,theory_mbl_long_range5,
theory_mbl_long_range6,arti_garg_2019,theory_fully_connected1,
theory_fully_connected2,theory_fully_connected3} have revealed the surprising fact that correlations in long-range hopping can actually aid localization \cite{theory_power_law_loc1,theory_power_law_loc3}. Several recent works investigate interacting systems with long-range hopping with the focus on the existence of many-body localization and entanglement in such systems \cite{theory_mbl_long_range1,theory_mbl_long_range2,theory_mbl_long_range3,
theory_mbl_long_range4,theory_mbl_long_range5,theory_mbl_long_range6,arti_garg_2019,Ranjan_Tanay,theory_entanglement1,
theory_entanglement2,theory_entanglement3,theory_entanglement4,theory_entanglement5,
theory_entanglement6,theory_entanglement7}.  There has also been a few recent works inspecting the physics of power-law hopping in presence of quasi-periodic potentials\cite{theory_power_law_loc4,theory_power_law_loc5,theory_power_law_quasi_periodic,Ranjan_Tanay}. Extremely rich phase-diagrams of such systems in terms of localization, delocalization and multifractality of the single-particle eigenfunctions have been presented \cite{theory_power_law_loc4, theory_power_law_loc5}.    

However, the localized states of disordered or quasi-periodic systems with power-law hopping are very different from those with short-ranged hopping. The localized states in presence of power-law hopping have a pronounced peak with tails decaying algebraically, instead of exponentially \cite{theory_power_law_loc1}. As a consequence, a single-particle localization length, if defined, would be infinite. In this sense, the algebraically localized states are not truly `localized'. So, unlike exponentially localized states, the relation between such algebraically localized states and the transport properties of the system in the thermodynamic limit is not obvious. Exploration of this physics is especially crucial in the context of all the recent works investigating many-body localization in long-range systems \cite{arti_garg_2019,theory_mbl_long_range1,theory_mbl_long_range2,theory_mbl_long_range3,
theory_mbl_long_range4,theory_mbl_long_range5,theory_mbl_long_range6}. But, to our knowledge, this has not been explored before. In this paper, we fill this gap by elucidating the connection between localization and transport for algebraically localized states. We show that in quasiperiodic systems with power-law hopping, the algebraically localized states can be actually conducting.

The most common characterization of a single-particle eigenstate as localized or delocalized is done in terms of the scaling of the Inverse Participation Ratio ($IPR$) with system-size\cite{theory_power_law_loc6,theory_power_law_loc4,theory_power_law_loc5}. For a localized state, the $IPR$ does not scale with system size. This property comes from the existence of a pronounced peak and is true both for exponentially localized and algebraically localized states. Indeed, in most of the previous works \cite{theory_power_law_loc6,theory_power_law_loc4,theory_power_law_loc5}, $IPR$ is one of the main quantities used to explore localization-delocalization transitions  is presence of power-law hopping. This, however, does not directly say anything about the transport properties of the system, as we show in this paper. We first prove that, for 1D non-interacting fermionic system at zero temperature, if the mean position of a particle at the Fermi energy is well-defined, then the system is insulating. If the corresponding mean position is ill-defined, then the system is conducting. According to this criterion, if the states near Fermi level are algebraically localized, the system can  be conducting or insulating, depending on the strength of the algebraic decay. We check this by working out an explicit example. The example we consider is the AAH model in presence of power-law hopping. We choose the strength of the AAH potential such that without power-law hopping, all states would be exponentially localized. As shown in Ref.\cite{theory_power_law_loc5}, for such choice of parameters, in presence of power-law hopping, there is a mobility edge in this system separating completely delocalized and algebraically localized states.  We consider $\alpha>1$, so that the thermodynamic limit is well-defined. We classify transport in terms of the Drude weight \cite{drude1,drude2,drude3} and the many-particle localization length \cite{many_particle4,many_particle5,many_particle3,many_particle2} (which is different from single-particle localization length) at zero temperature. We find that, when the Fermi level corresponds to an algebraically localized state, the system is conducting for $1<\alpha\leq 2$, while it is insulating for $\alpha>2$. We further show that due to quasi-periodicity of the AAH potential, there occurs near-resonance conditions, as a result of which, the mean of the probability distributions associated with the algebraically localized states become ill-defined for $1<\alpha\leq 2$, while the mean remains well-defined for $\alpha>2$. This thus exemplifies our analytical result. Note that this occurs due to quasi-periodic nature of the potential and will not be seen in case of random disorder. Most interestingly, for $1<\alpha<2$, we show clear evidence of super-diffusive transport through the algebraically localized states. Thus, we find a phase of the system, where there is a mobility edge in energy separating two different kinds of conducting regions, ballistic and super-diffusive. To our knowledge, this is the first time such a system is being reported.

\subsection{Localization and transport}\label{theory}
 A general Hamiltonian of a non-interacting a system with time-reversal symmetry is given by
\begin{align}
\hat{\mathcal{H}}=\sum_{\ell,m=-\ceil{N/2}}^{\ceil{N/2}} \mathbf{H}_{\ell m} \hat{c}_\ell^\dagger\hat{c}_m, 
\end{align}
where $\hat{c}_\ell$ is the bosonic or fermionic annihilation operator at site $\ell$, and $\mathbf{H}$ is a symmetric matrix (for time-reversal symmetry, $\mathbf{H}$ has to be real). For concreteness, we have numbered the sites from $-\ceil{N/2}$ to $\ceil{N/2}$, where $\ceil{x}$ the least integer greater than or equal to $x$. Finally, we will be interested in results in the thermodynamic limit, $N\rightarrow \infty$, where the exact numbering in the sum will not matter as long as the lower limit goes to $-\infty$ and the upper limit goes to $\infty$. The diagonal elements of $\mathbf{H}$ give the on-site energies, and the off-diagonal elements give the hopping, which in general can be long-range. The matrix $\mathbf{H}$ can be diagonalized via an orthogonal transformation
\begin{align}
\Phi^{T}\mathbf{H}\Phi = \mathbf{D},
\end{align} 
where $\mathbf{D}$ is a diagonal matrix containing the eigenvalues of $\mathbf{H}$, and $\Phi^{T}$ is the transpose of $\Phi$. The eigenvalues of $\mathbf{H}$ are the single particle energies of the system, and the columns of $\Phi$ give the single-particle eigenfunctions of the system. Localization phenomena in non-interacting systems concerns the localization of the single particle eigenstates. In the following, we restrict to 1D systems. 

Let $\Phi_{n}(x)$ be the single particle eigenstate of the system with energy $\omega_n$. Then, 
\begin{align}
P_n(x)=\Phi_{n}^2(x),
\end{align}
gives the probability of a particle to be found with energy $\omega_n$ at site $x$. The following defines a completely delocalized state,
\begin{align}
& P_n(x)\sim \frac{1}{N}~~\rightarrow\textrm{completely delocalized}.
\end{align}
 The single-particle state is termed localized if $P_n(x)$ has a pronounced peak at some cite, say $x_0$, with decaying tails. Depending on the nature of decay of the tails, the single-particle state can be exponentially localized or algebraically localized,
\begin{align}
\label{def_localization}
& P_n(x) \sim e^{-\left(\frac{|x-x_0|}{\zeta}\right)^p} ~~\rightarrow\textrm{exponentially localized} \\
& P_n(x) \sim \frac{1}{|x-x_0|^p},~|x-x_0|\gg 1\rightarrow\textrm{algebraically localized.}  \nonumber
\end{align} 
The IPR of the single-particle state is given by \cite{theory_power_law_loc4}
\begin{align}
IPR(n)=\sum_{x=-\ceil{N/2}}^{\ceil{N/2}} \Phi_{n}^4(x). 
\end{align} 
It can be readily checked that if the single-particle state is completely delocalized, i.e, $\Phi_{n}(x)\sim 1/\sqrt{N}$, the $IPR(n) \sim 1/N$. On the other hand, if the single-particle state has a pronounced peak that does not scale with system-size, $IPR(n) \sim N^0$, i.e, the $IPR$ does not scale with system-size. This property of $IPR$ holds irrespective of whether the single-particle state is algebraically localized or exponentially localized. Thus,  $IPR(n) \sim N^0$ is often taken as a defining property of a localized single-particle state. While $IPR$ does not distinguish between algebraically localized and exponentially localized states, various other related properties, like a full multifractal analysis, can be used to distinguish between algebraically localized and exponentially localized states where direct demonstration of the defining behavior in Eq.~\ref{def_localization} is not easy to obtain. The theory of localization is quite well-developed. But the connection between localization of single-particle eigenstates and particle transport properties of the system is non-trivial, and has not been explored for algebraically localized states. Exploring this connection is the goal of this work.

Particle transport properties of an isolated system in the thermodynamic limit is given by the Kubo formula.  At finite frequency, this is given by 
\begin{align}
\label{kubo_conductivity}
\sigma(\omega)= \pi D \delta(\omega) + \sigma^{reg}(\omega).
\end{align}
DC transport properties are given by the zero frequency limit of above formula. Here, $D$ is the Drude weight. The Drude weight gives the zero frequency peak of conductivity. A finite value of $D$ points to ballistic transport. DC conductivity diverges in such case. If transport in not ballistic, $D$ is zero. The second part, $\sigma^{reg}(\omega)$, which gives the regular part of conductance, governs transport properties in such cases.  If $\lim_{\omega\rightarrow0}\sigma^{reg}(\omega)=0$, the system is insulating. If $\lim_{\omega\rightarrow0}\sigma^{reg}(\omega)$ is finite, the system has normal diffusive transport, while if $\lim_{\omega\rightarrow0}\sigma^{reg}(\omega)\rightarrow\infty$, the conductivity diverges, even if $D$ is zero. This kind of transport is called super-diffusive.  

Eq.~\ref{kubo_conductivity} is strictly valid in the thermodynamic limit. For numerical calculations on finite systems, one has to be very careful of boundary conditions. It can be shown that for a finite system with open boundary conditions, the Drude weight $D$ is identically zero. This holds true even for ballistic transport, when periodic boundary conditions give a finite value of $D$. As shown in \cite{Rigol_drude}, in such cases, under open boundary conditions, $\sigma^{reg}(\omega)$ develops a peak at finite frequency, which grows in height and moves towards zero frequency as system-size is increased. So, in the thermodynamic limit, equivalence between open boundary and periodic boundary conditions is restored.

Let us first write down the expression for $\sigma(\omega)$ for 1D systems assuming open boundary conditions (see Appendix \ref{Kubo_deriv} for the derivation)
\begin{align}
\label{Kubo_obc}
\sigma(\omega)=&i\pi\int_{-\infty}^{\infty} dt e^{i\omega t} \nonumber\\
&\frac{d}{dt}\left( \lim_{N\rightarrow\infty} \frac{1}{N}\sum_{p,q=-\ceil{N/2}}^{\ceil{N/2}} pq \langle [\hat{n}_p(t), \hat{n}_q(0)] \rangle\right).
\end{align}
Here we have used the definition of particle current operator for open boundary condition: $\hat{I}=\frac{d}{dt}\hat{x}=\frac{d}{dt}\sum_{p=1}^Np\hat{n}_p$, and $\langle ... \rangle= Tr(e^{-\beta \hat{\mathcal{H}}}/{Z}...)$, with $\beta$ being the inverse temperature.  This is one of the so called `Einstein relations' connecting particle conductivity to `diffusion' of density correlations. In fact, the long time scaling of the density correlations is one of the standard ways to classify DC transport. After the Fourier transform in above equation, this becomes the low frequency scaling of $\sigma(\omega)$,  
\begin{align}
\label{low_freq_ansatz}
&\sigma(\omega) \sim \omega^{-s},~~\omega \rightarrow 0.
\end{align}
For ballistic transport $s=1$, for diffusive transport $s=0$, for super-diffusive transport $0<s<1$, for sub-diffusive transport $-1<s<0$, for complete absence of diffusion (henceforth we will identify this with the exponentially localized case) $s\leq -1$ (see Appendix~\ref{classification} for more detailed discussion of transport classification). This is true in general, for both interacting and non-interacting systems at all temperatures. We now specialize to non-interacting fermionic systems at zero temperature. For non-interacting systems, we can now evaluate the above expression in terms of single-particle eigenstates. At zero temperature, this is given by, 
\begin{align}
\label{Kubo_non_int}
&\sigma(\omega)=\lim_{N\rightarrow\infty} \frac{\pi}{N} \sum_{p,q=-\ceil{N/2}}^{\ceil{N/2}} pq  \sum_{\substack{m,n \\ \omega_m\leq E_F\\\omega_n\geq E_F}} \Big[  \nonumber \\ 
&(\omega_n-\omega_m)\Phi_n(p) \Phi_m(p)\Phi_n(q) \Phi_m(q)\delta(\omega+\omega_m-\omega_n)\Big].
\end{align}
From the above equation, it is clear that the low frequency behavior is governed by nature of single-particle eigenstates near the Fermi energy $E_F$. Thus, unsurprisingly, the nature of single-particle eigenstates near $E_F$ governs the DC transport properties of a non-interacting fermionic system in the thermodynamic limit at zero temperature. However, it is non-trivial to take the $\omega\rightarrow 0$ limit in the above equation, because it has to be taken only after taking the $N\rightarrow \infty$ limit. Otherwise the result is always zero. The physical reason for this is that in a finite system, there is always a infrared cut-off, given by $\omega_{min}\sim 1/N$. One way of consistently taking the low-frequency and the thermodynamic limit of the above equation is evaluating the summation at $\omega_{min}$ and approximating the $\delta$-function with a Lorentizian (any other approximation to $\delta$-function will also work) whose width is also $\omega_{min}$,
\begin{align}
\label{Kubo_Lorentzian}
&\lim_{\omega\rightarrow0}\sigma(\omega)=\lim_{\omega_{min}\rightarrow 0}\lim_{N\rightarrow\infty} \frac{1}{N} \sum_{p,q=-\ceil{N/2}}^{\ceil{N/2}} pq  \sum_{\substack{m,n \\ \omega_m\leq E_F\\\omega_n \geq E_F}} \Big[  \nonumber \\ 
& \frac{(\omega_n-\omega_m)\Phi_n(p) \Phi_m(p)\Phi_n(q) \Phi_m(q)~\omega_{min}}{(\omega_{min}+\omega_m-\omega_n)^2+\omega_{min}^2}\Big].
\end{align}
Now, if we take $\omega_{min}\rightarrow 0$ first, the term in the square bracket gives a finite value.  Assuming that there is no degeneracy of single-particle eigenvalues (for finite system under open boundary conditions), this yields
\begin{align}
\label{Kubo_and_mean}
\lim_{\omega\rightarrow0}\sigma(\omega) = \lim_{N\rightarrow\infty} \frac{(\overline{x}(E_F))^2}{N} ,
\end{align}
where
\begin{align}
\overline{x}(E_F)=\sum_{x=-\ceil{N/2}}^{\ceil{N/2}} x P_{n}(x)|_{\omega_n\approx E_F},
\end{align}
is the mean of the probability distribution given by square of single-particle eigenfunction at $E_F$. It relates conductivity of a non-interacting system at zero temperature to the mean position of a particle with energy $E_F$. In going from Eq.~\ref{Kubo_Lorentzian} to Eq.~\ref{Kubo_and_mean}, we have switched the order of taking the two limits. This is only allowed if the RHS of Eq.~\ref{Kubo_and_mean} is well-defined. However, the RHS may be ill-defined because the $N\rightarrow\infty$ limit of the mean of the probability distribution may be ill-defined.  The mean is well-defined in thermodynamic limit only if
\begin{align}
\label{mean_existence}
\lim_{N\rightarrow\infty}\overline{x}(E_F)&=\lim_{a\rightarrow\infty}\Big[\lim_{b\rightarrow \infty}\sum_{x=-a}^{b} x P_{n}(x)|_{\omega_n\approx E_F}\Big] \nonumber \\
&=\lim_{b\rightarrow \infty}\Big[\lim_{a\rightarrow\infty}\sum_{x=-a}^{b} x P_{n}(x)|_{\omega_n\approx E_F}\Big].
\end{align} 
In other words, the mean should be the same finite number irrespective of the order in which the lower limit is taken to $-\infty$ and the upper limit is taken to $\infty$. Each of the limits in above equation must exist. 
 It can be checked that the existence of the mean in the thermodynamic limit is the necessary and sufficient condition for the RHS of Eq.~\ref{Kubo_and_mean} to be well-defined (see Appendix \ref{RHS_existence}). Eq.~\ref{Kubo_and_mean} then shows that, if the mean is well-defined, $\lim_{\omega\rightarrow0}\sigma(\omega)=0$.
So, existence of $\overline{x}(E_F)$ in the thermodynamic limit is a sufficient condition for the system to be insulating at zero temperature. It can be checked that this is also a necessary condition for the same. To see this, note that naively setting $\omega=0$ in Eq.~\ref{Kubo_non_int}, gives $\sigma(0)=0$, which means, if the system is insulating, interchanging the order of the $\omega\rightarrow 0$ and the  $N\rightarrow \infty$ limits must be possible. Thus, Eq.~\ref{Kubo_and_mean} must be well-defined in such case, which implies that  $\overline{x}(E_F)$ is well-defined in the thermodynamic limit. \emph{ Thus, if and only if the mean position of a particle with Fermi energy is well-defined in the thermodynamic limit, the system is insulating. Conversely, it follows that, if the corresponding mean is ill-defined in the thermodynamic limit, then the system is conducting.} In this case, the RHS of Eq.~\ref{Kubo_and_mean} is ill-defined which shows that switching of the order of the limits in Eq.~\ref{Kubo_Lorentzian} is not allowed. Hence we have proven the following crucial result: for 1D non-interacting fermionic systems at zero temperature,
\begin{align}
\label{loc_criterion}
&\lim_{N\rightarrow\infty}\overline{x}(E_F)\rightarrow\textrm{well-defined}\Rightarrow\textrm{insulating} \nonumber \\
&\lim_{N\rightarrow\infty}\overline{x}(E_F)\rightarrow\textrm{ill-defined}\Rightarrow\textrm{conducting},
\end{align} 
assuming no degeneracy of single-particle eigenvalues for the finite system under open boundary conditions (which is true in most cases).
It directly connects properties of single-particle eigenfunctions in 1D non-interacting systems to transport. Let us understand this result in terms of localization. Intuitively, spatial localization of a single particle eigenstate means that one can associate a `classical' notion of position to a particle in that state. This is certainly true for an exponentially localized state, where the particle can be taken to located in a region of width $\zeta$ around $x_0$ (see Eq.~\ref{def_localization}). Here $\zeta$ is the single particle localization length. In this case, the mean position is certainly well-defined, and thereby from Eq.~\ref{loc_criterion}, the system is insulating if the Fermi level corresponds to an exponentially state. This is also intuitive since the particle can be taken to be confined to a finite region. The situation is more complicated if there is an algebraically localized state at $E_F$. In this case, there is no length scale within which the particle can be assumed to be confined. In other words, an infinite number of moments of the probability distribution $P_n(x)$ are ill-defined in the thermodynamic limit. Eq.~\ref{loc_criterion} then gives the extremely non-trivial result that it is the existence of the mean (i.e, the first moment) of $P_n(x)$ that governs the nature of transport. If the mean exists, then the system is insulating, irrespective of non-existence of higher moments. Conversely, if the mean is ill-defined, there is no way to associate `classical' notion of position to a particle in that state (i.e., not even a mean position). \emph{All} moments of $P_n(x)$ are ill-defined in this case.   A particle in such a state cannot really be considered `spatially localized' but must be considered extended over the entire system. Since the state is extended (in this sense), it should contribute to transport.

Eq.~\ref{loc_criterion} thus has profound consequences for the algebraically localized states. It is well-known that mean of a probability distribution with tails decaying as $P_n(x)\sim |x-x_0|^{-p}$ is well-defined only if $p>2$. This can be easily checked by noting that for $P_n(x)\sim |x-x_0|^{-p}$, $|x-x_0|\gg 1$
\begin{align}
\sum_{x=a}^N x P_n(x) \sim 
\begin{cases}
&N^{2-p},~\forall~p\neq 2 \\
&\log(N),~ p=2,
\end{cases}
\end{align}
for arbitrary finite choice of  $a$. Thus, for $p\leq 2$, one of the limits in Eq.~\ref{mean_existence} does not exist, while for $p>2$ all the limits in Eq.~\ref{mean_existence} can be shown to exist. By Eq.~\ref{loc_criterion}, this says that, 
\begin{align}
\label{alg_loc_criterion}
& \textrm{For } P_n(x)|_{\omega_n\approx E_F}\sim \frac{1}{|x-x_0|^p}, |x-x_0|\gg 1, \nonumber \\
&\begin{cases}
& p>2, \Rightarrow\textrm{insulating}\\
&p\leq 2, \Rightarrow\textrm{conducting}.
\end{cases} 
\end{align}
Thus, we have found the connection between algebraic localization in 1D and transport. This is our main analytical result. Most interestingly, this says that, contrary to exponential localization, algebraically localized states can be conducting. Note that, $IPR\sim N^0$ for any positive the value of $p$. \emph{So for $p\leq 2$, we have states which are localized according to $IPR$ scaling with system-size, but are conducting in the thermodynamic limit.}

We will like to check the above findings in a non-trivial setting. For this purpose, we have to characterize DC transport at zero temperature numerically. DC transport is characterized by the low-frequency behavior of $\sigma(\omega)$ (see Eq.~\ref{low_freq_ansatz}), and hence by the existence of the Drude weight and by the low-frequency nature of $\sigma_{reg}(\omega)$. However, direct numerical calculation of these quantities for large system-sizes is quite difficult. At zero temperature, there exists an alternative, easier but equivalent way which we discuss below.   

At zero temperature under periodic boundary conditions, it was shown by Kohn \cite{drude1} that the Drude weight can be equivalently calculated from the change in the ground state energy of the system in presence of a small magnetic flux. Let $E_0$ be the ground state energy of the system in presence of a flux $\phi$. Then, the Drude weight is given by
\begin{align}
\label{D_def}
D(N) = \frac{N}{4\pi^2}\frac{\partial^2 E_0}{\partial\phi^2}|_{\phi\to\phi_{min}},~~D=2\pi\lim_{N\rightarrow\infty} D(N)
\end{align}
where $\phi_{min}$ is the flux at which $E_0$ becomes minimum. For a non-interacting system, $D(N)$ is governed by the nature of the single-particle eigenstates of the system near the Fermi energy $E_F$ (since $\lim_{\omega\rightarrow 0}\sigma(\omega)$ is governed by the same, see Eq.~\ref{Kubo_non_int}). From Eq.~\ref{D_def}, we see that $D(N)$ corresponds to the change in the ground state energy of the system with periodic boundary condition under an infinitesimal flux. Since putting a flux corresponds to a twist in the boundary conditions, $D(N)$ measures the change in $E_0$ due to a small change in boundary conditions. Thus, it is plausible that finite-size scaling of $D(N)$ depends on the weight of the eigenfunctions at the boundary.   If the states near the Fermi energy $E_F$ are completely delocalized, which corresponds to ballistic transport, the weight of eigenfunctions at the boundary does not decay with system-size. So $D(N)\sim N^0$, which corresponds to ballistic transport. If the states near $E_F$ are exponentially localized, the weight of the eigenfunctions at the boundary decay exponentially, and $D(N)\sim e^{-N}$.  By exact same reasoning, if the states near $E_F$ are `algebraically localized', we expect $D(N)$ to decay as a power-law, $D(N)\sim N^{-\ell}$.

Thus, if $D(N)\sim N^0$, transport is ballistic. For an exponentially localized system, $D(N)\sim e^{-N}$. For other types of transport (diffusive, super-diffusive, sub-diffusive),  $D(N)$ goes to zero with system size slower than exponentially. For algebraically localized states, we expect, $D(N)\sim N^{-\ell}$.

Further classification is provided by $\lim_{\omega\rightarrow0}\sigma^{reg}(\omega)$. But, direct calculation of this limit of $\sigma^{reg}(\omega)$ is difficult. So following refs.~\cite{many_particle4,many_particle5,many_particle3,many_particle2} one can equivalently look at the many-particle localization length, which is defined as follows.
Let $|0\rangle$ be the many-particle ground state of the system,
\begin{align}
|0\rangle=\sum_{n=1}^{N_e} |\Phi_n\rangle.
\end{align} 
Here, $N_e$ is the number of particles in the system. The many-particle localization length $\xi$ is defined as \cite{many_particle4,many_particle5,many_particle3,many_particle2} 
\begin{align}
\xi^2(N) &= \frac{1}{N} \left(\langle 0 | \hat{x}^2 | 0 \rangle - \left(\langle 0 | \hat{x} | 0 \rangle\right)^2 \right) \nonumber \\
& = \frac{1}{2N} \sum_{p,q=-\ceil{N/2}}^{\ceil{N/2}} \left[ (p-q)^2 \left( \langle \hat{n}_p \hat{n}_q \rangle - \langle \hat{n}_p \rangle \langle \hat{n}_q \rangle \right)\right], 
\end{align} 
where $\hat{x}=\sum_{p=-\ceil{N/2}}^{\ceil{N/2}} p\hat{n}_p$ is the position operator, and $\langle ... \rangle = Tr(|0\rangle\langle 0|...)$. (In Refs.~\cite{many_particle4,many_particle5,many_particle3,many_particle2}, the definition of $\xi^2$ involves normalization by number of particles $N_e$. Here we have instead normalized by $N$ assuming $N_e\propto N$. This does not change any of the physics associated with $\xi^2(N)$.)  Once again, it is important to note the issue of boundary conditions. The many-particle localization length requires definition of the position operator. The position operator is well-defined only in open boundary conditions. So, contrary to the Drude weight, for numerical calculations on a finite-size system, one needs to calculate $\xi^2(N)$ strictly under open boundary conditions.  The many-particle localization length helps us to characterize transport due to the following relation valid at zero temperature (see Refs.~\cite{many_particle4,many_particle5,many_particle3,many_particle2} and also Appendix \ref{xi_Kubo_reg} for the derivation),
\begin{align}
\label{def_xi}
\lim_{N\rightarrow\infty}\xi^2(N)  \propto \lim_{N\rightarrow\infty}\int_{\omega_{min}}^{\omega_{max}} d\omega \frac{\sigma^{reg}(\omega)}{\omega},
\end{align}
with $\omega_{min}\sim 1/N$ and $\omega_{max}\sim$ inverse of lattice-spacing. In above equation, $\sigma^{reg}(\omega)$ is assumed to be calculated in a finite but large system with open boundary conditions, and then the $N\rightarrow\infty$ limit is taken in the RHS.   Note that on a finite but large system with open boundary conditions, $\sigma^{reg}(\omega_{min})\sim \sigma(\omega_{min})$.   Hence using Eq.~\ref{low_freq_ansatz} in Eq.~\ref{def_xi}, and noting $\omega_{min}\sim 1/N$, we can find the behavior of $\xi^2(N)$ with system-size as
\begin{align}
\xi^2(N) \sim
\begin{cases}
&N^{s},~~\forall~ s>0 \\
&log(N),~~\forall~s=0 \\
& N^0,~~\forall~s<0
\end{cases},
\end{align}
So we see that $\xi^2(N)$ is finite in thermodynamic limit for sub-diffusive transport, $s<0$, which gives insulating behavior in thermodynamic limit.  So, finiteness of $\xi^2(N)$ points to insulation. On the other hand, logarithmic divergence of  $\xi^2(N)$  shows the system is diffusive, and has a finite conductivity. Power-law divergence of $\xi^2(N)$ shows conductivity is diverging. For ballistic transport, $\xi^2(N)\sim N$. Like $D(N)$, for non-interacting systems, $\xi^2(N)$ is governed by the nature of states near $E_F$ (since $\lim_{\omega\rightarrow 0}\sigma(\omega)$ is governed by the same, see Eq.~\ref{Kubo_non_int}).

So, finite size scaling of $D(N)$ under periodic boundary conditions and that of $\xi^2(N)$ under open boundary conditions allow us to characterize zero temperature particle transport as follows,
\begin{align}
\label{transport_classification}
& D(N)\sim N^0, ~ \xi^2(N) \sim N~\Rightarrow\textrm{conducting, ballistic,} \nonumber \\
& D(N)=~N^{-\ell}, ~ \xi^2(N) \sim N^{s}~\Rightarrow\textrm{conducting,} \nonumber\\
& \hspace{124pt} \textrm{super-diffusive,} \\
& D(N)=~N^{-\ell}, ~ \xi^2(N) \sim \log(N)~\Rightarrow\textrm{conducting, diffusive,} \nonumber \\
& D(N)=~N^{-\ell}, ~ \xi^2(N) \sim N^0~\Rightarrow\textrm{insulating, sub-diffusive,} \nonumber \\
& D(N)\sim e^{-N}, ~ \xi^2(N) \sim N^{0}~\Rightarrow\textrm{insulating, exponentailly} \nonumber \\ 
& \hspace{124pt} \textrm{localized,} \nonumber
\end{align}
where we have assumed that the slower than exponential decay of $D(N)$ with $N$ for the middle three cases is a power-law decay. This, as explained before, is expected for the algebraically localized systems.  Note that both $D(N)$ and $ \xi^2(N)$ are properties of the many-particle ground state of the system, and are well-defined irrespective of whether the system is interacting or non-interacting. Here, however, we are interested in non-interacting systems, where nature of single-particle eigenstates near $E_F$ govern the behavior of $D(N)$ and $ \xi^2(N)$. Particularly, we see from Eq.~\ref{alg_loc_criterion}, if the states near $E_F$ are algebraically localized, then, for $p>2$, the system will be insulating and $ \xi^2(N)$ will remain finite in the thermodynamic limit, while for $p\leq 2$, the system is expected to be conducting, and $ \xi^2(N)$ with diverge with increase in system size. In either case, $D(N)$ is expected to go to zero as a power-law.  In the following, we work out an illustrative example where exactly this happens. To our knowledge, this is the first work showing that algebraically localized states in 1D can be conducting.

\subsection{An illustrative example}\label{numerical_example}
\subsubsection{The model}

We consider a model with the quasiperiodic Aubry-Andr{\'e}-Harper potential and power-law hopping 
\begin{align}
\label{AAHlr}
&\hat{H}=\sum_{x=-\ceil{N/2}}^{\ceil{N/2}} \varepsilon(x) \hat{c}_x^{\dagger} \hat{c}_x \nonumber \\
&-\sum_{x=-N/2}^{N/2} \sum_{m=1}^{\ceil{N/2}-1} \left(\frac{1}{m^{\alpha}} \hat{c}^{\dagger}_{x} \hat{c}_{x+m} + h.c.\right), \nonumber \\
& \varepsilon(x) = W \cos(2\pi b~x) 
\end{align}
Here $\{c_x\}$ is the fermionic annihilation operator at site $x$, and $b$ is an irrational number. The system has power-law hopping with exponent $\alpha$ and strength $-1$.   The long-range power-law hopping has a hard cut-off at $m=\ceil{N/2}-1$.  This cut-off is required to uniquely define the periodic boundary condition, which is required for calculation of the Drude weight. The important point here is that the cut-off scales with system-size. In addition, the system has an on-site potential $\varepsilon(x)$, which is a cosine potential of strength $W$ and period $1/b$. If $b$ is an irrational number, the period of the potential is incommensurate with the lattice. We take $b=(\sqrt{5}-1)/2$, which is the golden mean. Rational approximations to the golden mean is given by the ratios of consecutive Fibonacci numbers,
\begin{align}
\label{def_Fibo_golden_mean}
F_n = F_{n-2}+F_{n-1},~\lim_{n\rightarrow \infty}\frac{F_n}{F_{n+1}}=b=\frac{\sqrt{5}-1}{2},
\end{align}
where $F_n$ is the $n$th Fibonacci number. To implement periodic boundary condition along with this incommensurate potential, the system-sizes are chosen to be Fibonacci numbers.  

\begin{figure*}
\includegraphics[width=\linewidth, height=9.5cm]{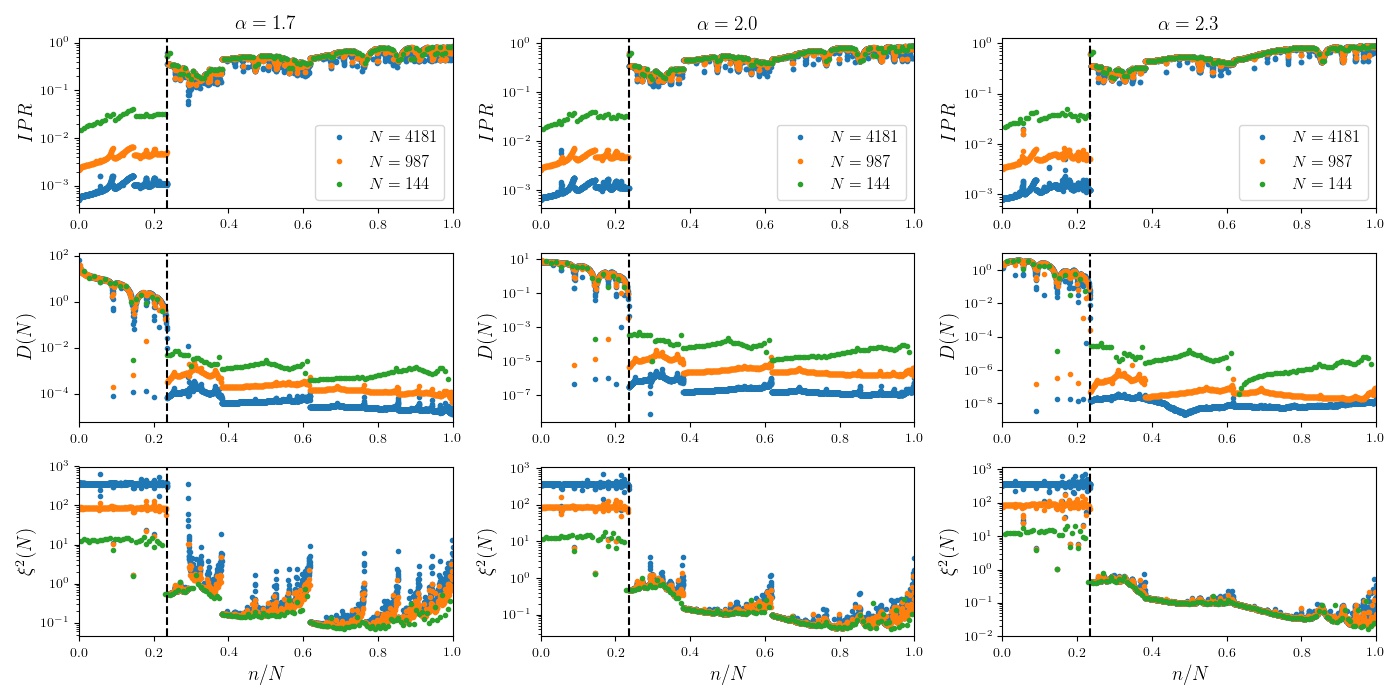}
\caption{(Color online) The figure shows plots of $IPR$, $D(N)$ and $\xi^2(N)$ for as a function of $n/N$ where $n$ is the single-particle eigenstate index. For $D(N)$ and $\xi^2(N)$, $n/N$ is to be interpreted as the ground state filling fraction, i.e, all single-particle eigenstates up to $n$th state are occupied and the rest are empty. The left column is for $\alpha=1.7$, which is representative of $\alpha<2$, the right column is for $\alpha=2.3$, which is representative of $\alpha>2$, the middle column is for $\alpha=2$. Each plot shows results for three different system-sizes. The vertical dashed line in all plots corresponds to $b^3$, which is the fraction of completely delocalized states. $W=3$.} 
\label{fig:IPR_Drude_mpll_3_values}
\end{figure*}

With nearest neighbour hopping (i.e, $\alpha\rightarrow\infty)$, the above model is the paradigmatic Aubry-Andr{\'e}-Harper model. This model shows a phase transition from all states completely delocalized to all states exponentially localized with increase in the strength of on-site potential $W$. The transition point is $W=2$, which is the critical point. There is no mobility edge in the nearest neighbour case. In presence of power-law hopping, there occurs a rich phase diagram of the model with mobility edges separating different kinds of states. The phase diagram of the model in terms of the mobility edges  has been explored in detail in a recent work \cite{theory_power_law_loc5}. One of the main results of that work is that, depending on the value of $W$ and $\alpha$ there occurs a fraction $b^q$ number of completely delocalized states, where $q$ is an integer. The parameter space in terms of $\alpha$ and $W$ can be broken up into regions with integer values of $q$. A single mobiltiy edge occurs separating these states from the rest of the states. For $\alpha<1$, the rest of the states are multifractal, while for $\alpha>1$, the rest of the states are algebraically localized.     

In this paper, we want to look at particle transport at zero temperature through a 
non-interacting system which has algebraically localized states near $E_F$. So, throughout the rest of the paper, for numerical calculations, we choose, $W=3$, and $\alpha>1$. The classification of transport behavior in terms of $D(N)$ and $\xi^2(N)$ given in the previous section depends on the existence of the thermodynamic limit. Since we are dealing with a long-range system the existence of the thermodynamic limit is not obvious. However, for $\alpha>1$, the single particle eigen-energies are bounded from below in the thermodynamic limit. Also, the ground state energy at a given filling is extensive, as can be checked by explicit numerical calculations (see Appendix~\ref{appendix:therm_limit}). So, in this case, the thermodynamic limit is well-defined.

\subsubsection{Numerical results}

\begin{figure*}
\includegraphics[width=\linewidth]{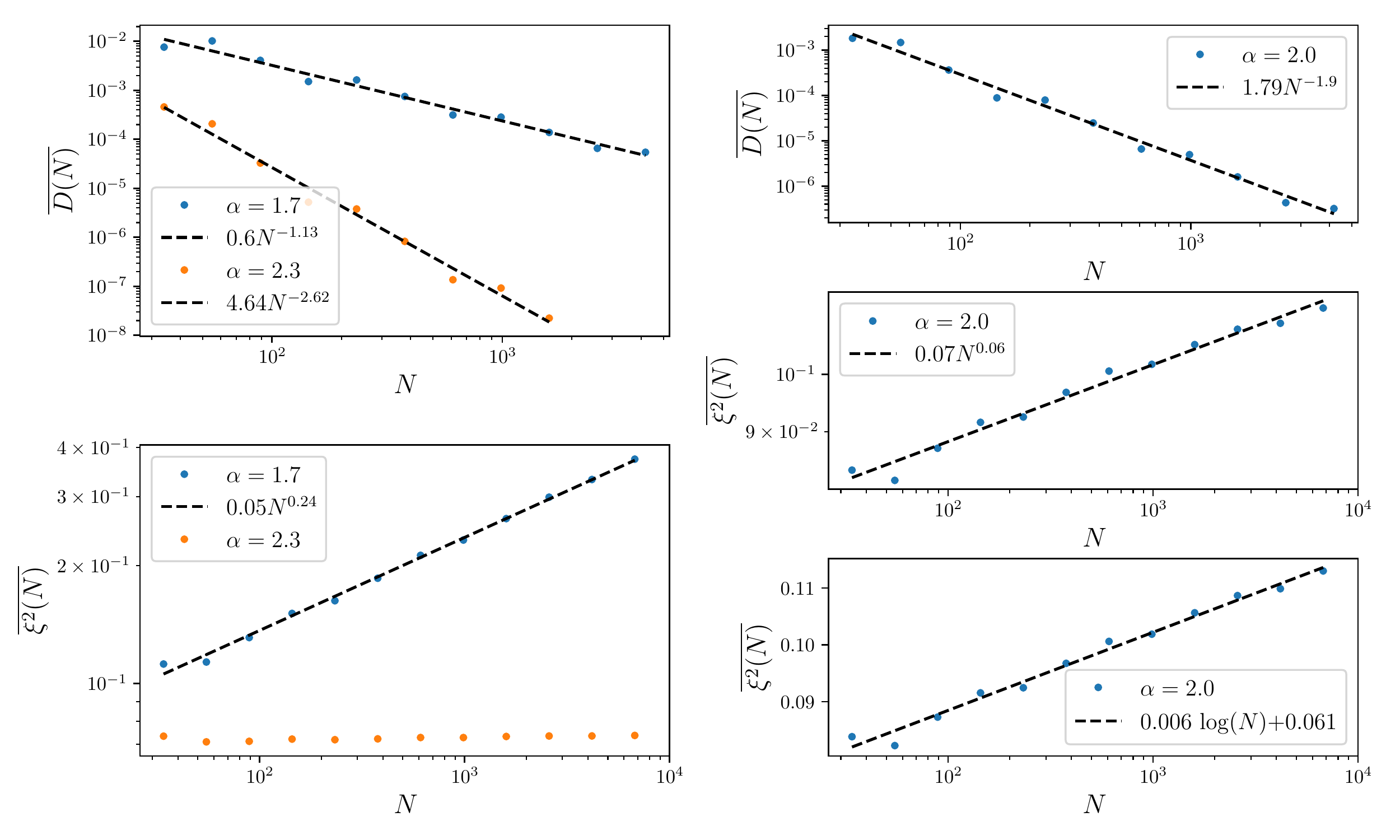}
\caption{(Color online) System-size scaling of $\overline{D(N)}$ and $\overline{\xi^2(N)}$, which are the the average values of $D(N)$ and $\xi^2(N)$, averaged over all fillings where the Fermi energy's correspond algebraically localized states. Left panel shows the plots of $\overline{D(N)}$ and $\overline{\xi^2(N)}$ with $N$ for two different values of $\alpha$, $\alpha=1.7$ and $\alpha=2.3$. The right panel shows plots of $\overline{D(N)}$ and $\overline{\xi^2(N)}$ with $N$ for $\alpha=2$. The middle plot of the right panel shows a log-log plot with a power-law fit of $\overline{\xi^2(N)}$ for $\alpha=2$. The bottom plot of the right panel shows a log-linear plot with a logarithmic fit of $\overline{\xi^2(N)}$ for $\alpha=2$.  $W=3$.} 
\label{fig:scaling_plots}
\end{figure*}

In 1D systems with long-range hopping, a lot of interesting effects are seen  when the hopping exponent is $1<\alpha<2$, which are often markedly different from $\alpha>2$. The interesting transport properties of the ordered system in this regime have been recently reported in \cite{ordered_transport} by the authors. In the context of quasi-periodic systems, some of the interesting features in this regime have been discussed in \cite{theory_power_law_loc5}. In the present case also, we will see that $1<\alpha<2$ and $\alpha>2$ will have markedly different behaviors. 

In Fig.~\ref{fig:IPR_Drude_mpll_3_values}, we present the numerical results for $IPR$, $D(N)$ and $\xi^2(N)$ for three values of $\alpha$: $\alpha=1.7$ (which is representative for $1<\alpha<2$), $\alpha=2$ and $\alpha=2.3$ (which is representative for $\alpha>2$). For these values of $\alpha$ and our chosen value of $W=3$, according to \cite{theory_power_law_loc5}, there are $b^3$ fraction of completely delocalized states and the rest of the states are algebraically localized (for explicit illustrative plots showing algebraically localized and exponentially localized states, refer to Appendix~\ref{appendix:alg_exp_loc}). In Fig.~\ref{fig:IPR_Drude_mpll_3_values}, all the quantities are plotted against $n/N$, where $n$ is the single-particle eigenstate index, with the single-particle eigenvalues arranged in ascending order. The points in the $IPR$ plots correspond to the $IPR$ of the single-particle eigenstates. For $D(N)$ and $\xi^2(N)$ plots, $n/N$ gives the ground state filling fraction. The vertical dashed lines in all the plots correspond to $b^3$. The fraction of states with $n/N<b^3$ are completely delocalized. So their $IPR\sim N^{-1}$ (as can be checked by multiplying the data points by $N$, see Appendix~\ref{appendix:scaling_deloc}), $D(N)\sim N^0$ (as can be seen from the plots), $\xi^2(N) \sim N$ (as can be seen by dividing the data points by $N$, see Appendix~\ref{appendix:scaling_deloc}). Our main object of interest is the typical behavior of the rest of the states, i.e, the states for which $n/N>b^3$. It is clear from the plots that for these states $IPR\sim N^0$, which clearly points towards localization. However, at these filling fractions, we see that $D(N)\sim N^{-\ell}$. This is consistent with our expectation for algebraically localized states, and confirms that the states are not exponentially localized. This is true for all values of $\alpha>1$.  Most interestingly, we see from the plots that $\xi^2(N)$ for filling fraction $n/N>b^3$ behave differently for $\alpha\leq 2$ and $\alpha>2$. For $\alpha\leq 2$, $\xi^2(N)$ seems to increase with system-size, while for $\alpha>2$, seems to not scale with system-size. As discussed before, this suggests that for  $\alpha\leq 2$, the algebraically localized states are conducting, while for $\alpha>2$, the algebraically localized states are insulating. Note that there is a more intricate structure and possible multiscaling, especially for $\alpha=2$. This is the usual case for quasi-periodic systems due to self-similar singular spectra of eigenenergies. While this is interesting, here, we will not be concerned with such details. Instead in the following, we will be looking at the behavior of system averaged over all values of $n$ with $n/N>b^3$.

We denote $\overline{D(N)}$ and $\overline{\xi^2(N)}$ by the values of $D(N)$ and $\xi^2(N)$ respectively, averaged over all filling fractions where the Fermi energy corresponds to an algebraically localized state, i.e., for $n/N>b^3$. Figure.~\ref{fig:scaling_plots} top left panel shows plots of $\overline{D(N)}$ with $N$ for $\alpha=1.7$ and $\alpha=2.3$. In both cases,  $\overline{D(N)}$ decays as a power-law. Figure.~\ref{fig:scaling_plots} bottom left panel shows plots of $\overline{\xi^2(N)}$ with $N$ for $\alpha=1.7$ and $\alpha=2.3$. Here, we see that for $\alpha=1.7$, $\overline{\xi^2(N)}$ diverges with $N$ as a power-law with an exponent between $0$ and $1$, while for  $\alpha=2.3$, $\overline{\xi^2(N)}$ does not scale with system-size. Thus, according to the classification of transport properties in Eq.~\ref{transport_classification}, the transport through algebraically localized states is super-diffusive for $\alpha=1.7$. We have checked that this is the case for $1<\alpha<2$. So, in this regime, the algebraically localized states are conducting, with a diverging conductivity. On the other hand, for $\alpha>2$, the transport through algebraically localized states is sub-diffusive according to Eq.~\ref{transport_classification}. In this regime, the algebraically localized states are insulating. 

\begin{figure}[t]
\includegraphics[width=\columnwidth]{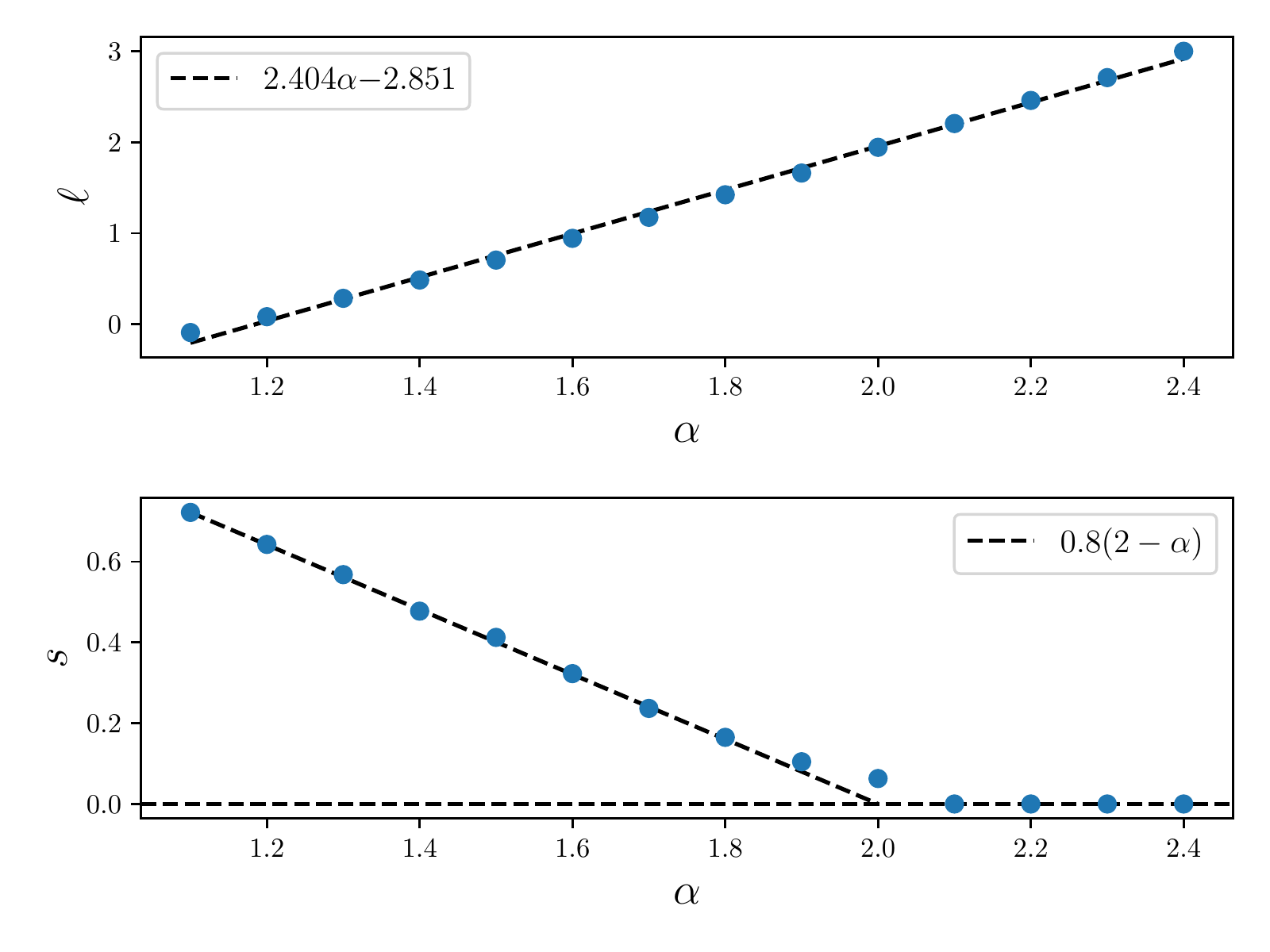}
\caption{(Color online) The figure shows plots of power-law scaling exponents $\ell$ and $s$, corresponding to $\overline{D(N)}\sim N^{-\ell}$ and $\overline{\xi^2(N)}\sim N^{s}$, with $\alpha$. The results are obtained from power-law fits.  $W=3$.} 
\label{fig:scaling_exponents_plots}
\end{figure}

Figure.~\ref{fig:scaling_plots} right panel shows plots of $\overline{D(N)}$ and $\overline{\xi^2(N)}$ with $N$ for $\alpha=2.0$. From the top right plot, it is clear that $\overline{D(N)}$ decays as a power-law in this case also. However, the scaling of $\overline{\xi^2(N)}$ with $N$ seems to match equally well both a power-law fit with a very small exponent (right middle panel) and a fit of logarithmic divergence (right bottom panel). From our data, it is not possible to differentiate between these two cases, so we cannot conclude whether the transport is diffusive or weakly super-diffusive. Nevertheless, it is clear that at $\alpha=2$, the algebraically localized states are conducting.

\begin{figure}[t]
\includegraphics[width=\columnwidth]{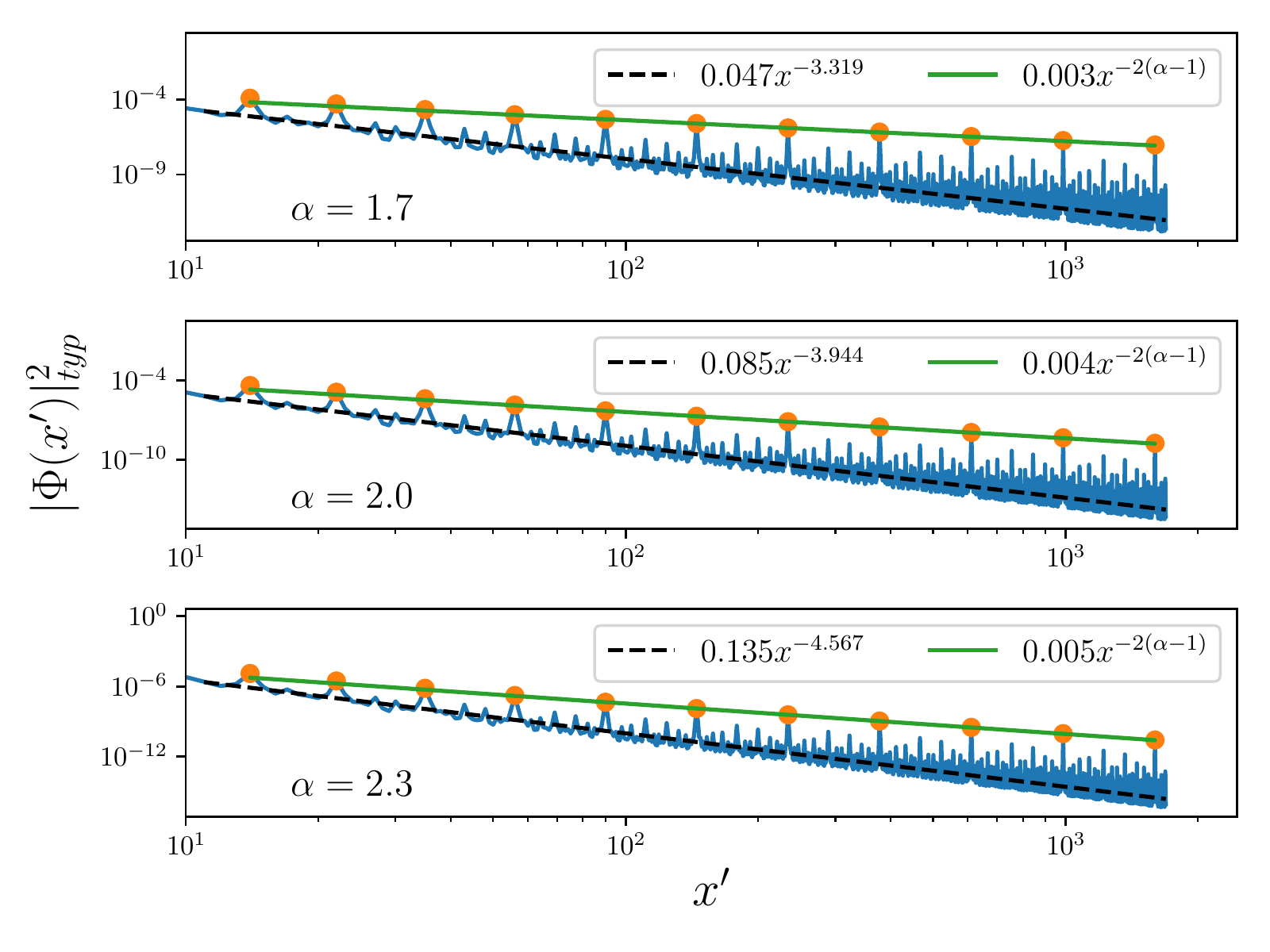}
\caption{(Color online) The figure shows scaling of $|\Phi_{n}(x^\prime)|^2_{typ}$ with $x^\prime$ for $\alpha=1.7$ (top), $\alpha=2.0$ (middle), $\alpha=2.3$ (bottom). The orange dots correspond to points where $x^\prime=$Fibonacci number. The green line shows the scaling of  $|\Phi_{n}(x^\prime)|^2_{typ}$ for these points. The black dashed line shows a least square fit of all the data points. System size: $N=6765$.  $W=3$.} 
\label{fig:tail_scaling}
\end{figure} 

Say $\overline{D(N)}\sim N^{-\ell}$ and $\overline{\xi^2(N)}\sim N^{s}$. In Fig.~\ref{fig:scaling_exponents_plots} we show the variation of $\ell$ and $s$ with $\alpha$ for $\alpha>1$. We see that $\ell$ increases linearly with $\alpha$. More interestingly, we observe that, for $1<\alpha<2$, $s\propto (2-\alpha)$. However, $\alpha=2$ gives a non-zero value of $s$ from a power-law fit. This seems to suggest that the behavior at $\alpha=2$ is indeed different from that for $1<\alpha<2$. This, though not at all conclusive, seems to point in favour of the logarithmic fit and hence, diffusive transport at $\alpha=2$.

Having established that the algebraically localized states are conducting on average for $\alpha\leq 2$, let us see if this is consistent with Eq.~\ref{alg_loc_criterion}. For this purpose, we look at the behavior of the power-law tails of the algebraically localized states. We denote by $\Phi_n(x)$, the single-particle eigenfunction of the $n$th eigenstate. The behavior of the power-law tails is embodied by the following quantity,
\begin{align}
\label{def_phi_typ}
&|\Phi_{n}(x^\prime)|^2_{typ} = \exp\left(\frac{1}{\mathcal{N}}\sum_{n}^\prime \log\left(|\Phi_{n}(x^\prime)|^2\right)\right), \\
& x^\prime = x-x_0,~~x>x_0, \nonumber
\end{align}
where $x_0$ is the position of the peak of the algebraically localized state, $\mathcal{N}$ is the number of algebraically localized states, $\sum_{n}^\prime$ denotes sum over all algebraically localized states. The above quantity is the geometric mean of absolute value square of all algebraically localized eigenfunctions, with the position of the peak shifted to zero (the physics discussed below holds even if arithmetic mean was taken instead of geometric mean, see Appendix~\ref{appendix_avg_phi}). The behavior of $|\Phi_{n}(x^\prime)|^2_{typ}$ with $x^\prime$  gives the typical decay of algebraically localized eigenfunctions, $|\Phi_{n}(x^\prime)|^2_{typ} \sim{ x^\prime}^{-p}$. From Eq.~\ref{alg_loc_criterion}, the value of $p$ governs whether the system is conducting or insulating.

The plots of $|\Phi_{n}(x^\prime)|^2_{typ}$ with $x^\prime$ are shown in Fig.~\ref{fig:tail_scaling} for three values of $\alpha$. For all values of $\alpha$ we see that a least square fitting gives a power-law decay with exponent $\sim 2\alpha$. However, more importantly, $|\Phi_{n}(x^\prime)|^2_{typ}$ shows a series of secondary peaks at values where $x^\prime$ is equal to a Fibonacci number. The height of these peaks decay as a power-law with the exponent given by $2(\alpha -1)$, i.e,
\begin{align}
|\Phi_{n}(F_n)|^2_{typ} &\sim F_{n}^{-2(\alpha-1)}.
\end{align}
 Then, from Eq.~\ref{alg_loc_criterion}, the system is insulating if 
\begin{align}
2(\alpha-1)>2 \Rightarrow \alpha>2.
\end{align}  
So, for $\alpha>2$, the system is insulating in thermodynamic limit. On the other hand, for $\alpha\leq 2$, from Eq.~\ref{alg_loc_criterion}, the system is conducting.
This is completely consistent with our findings from system-size scalings of $D(N)$ and $\xi^2(N)$. 

\begin{figure}[t]
\includegraphics[width=\columnwidth]{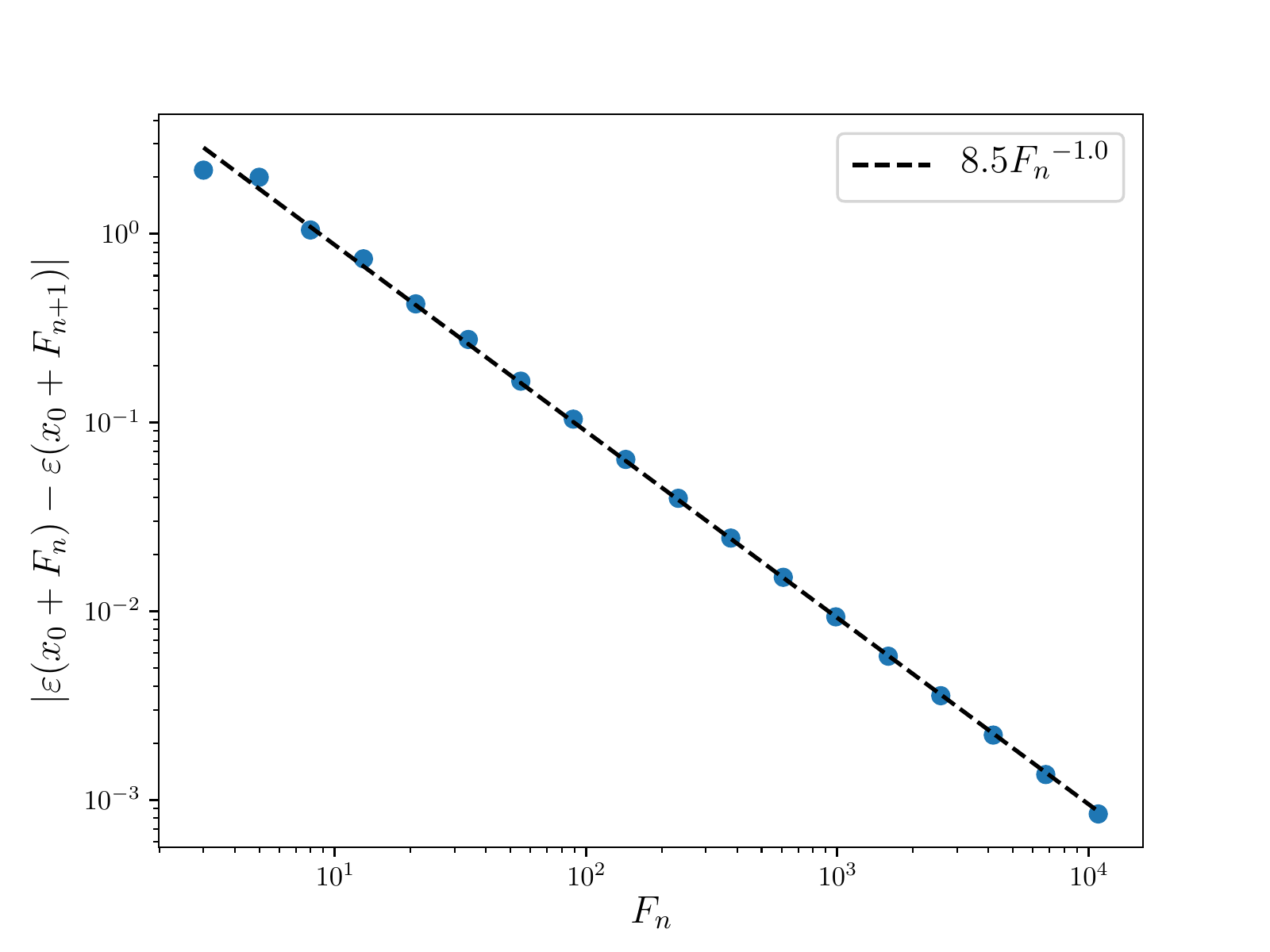}
\caption{(Color online) The figure shows that of  $|\varepsilon(x_0~+~F_n)~-~\varepsilon(x_0~+~F_{n-1})| $ decays as $1/F_n$. Here $x_0=1$. We have checked that the same remains true for any integer value of $x_0$. } 
\label{fig:AAH_pot_prop}
\end{figure} 

The above results show that the transport through the algebraically localized states in our example is completely governed by the occurrence of the secondary peaks at $|\Phi_{n}(F_n)|^2_{typ}$.  Let us now see the origin of these peaks. We note that because of the properties of Fibonacci numbers (Eq.~\ref{def_Fibo_golden_mean}), the on-site energies of sites separated by a distance equal to a Fibonacci number are very close to being in resonance. In particular, as shown in Fig.~\ref{fig:AAH_pot_prop} , we have,
\begin{align}
\label{AAHpot_prop}
|\varepsilon(x_0+F_n) - \varepsilon(x_0+F_{n-1})| \propto \frac{1}{F_{n}},
\end{align} 
for any integer value of $x_0$.
We also note that
\begin{align}
\label{Fibo_aymp_prop}
F_{n}-F_{n-1} = F_{n-2} \simeq b^2 F_{n},
\end{align} 
for large $n$. The decay of $|\Phi_{n}(x^\prime)|^2_{typ}$ with $x^\prime$ is governed by the degree of hybridization between the various sites of the system. Without the power-law hopping term, the system would be exponentially localized for our choice of parameters. So, the degree of hybridization between two far-off sites would be exponentially small. The algebraic decay of eigenfunctions is thus governed solely by the degree of hybridization between far-off sites due to the long-range hopping. Due to Eq.~\ref{AAHpot_prop}, the hybridization between sites separated by distances of Fibonacci numbers has a tendency to increase with the increasing value of the Fibonacci number, while due to the power-law decay of hopping, it has a tendency to decrease. The height of a secondary peak  occurring at a distance of $F_n$ from the main peak of an algebraically localized eigenfuction is governed by its degree of hybridization with the previous peak at $F_{n-1}$.  By this argument, we see that, in our case,
\begin{align}
|\Phi_{n}(F_n)|^2_{typ} &\sim \Big|\frac{1}{(F_n-F_{n-1})^{\alpha} (\varepsilon(F_n)-\varepsilon(F_{n-1}))}\Big|^2 \nonumber \\
& \sim F_{n}^{-2(\alpha-1)},
\end{align}
where, in the second line, we have used Eqs.~\ref{AAHpot_prop},~\ref{Fibo_aymp_prop}. This is exactly as we have seen from the numerical calculations in Fig.~\ref{fig:tail_scaling}.

Thus, we have shown, due to the quasi-periodic nature of the AAH potential, there occurs near resonance conditions, because of which, the algebraically localized states become conducting for $1<\alpha<2$. This is especially remarkable because usually mobility edges are thought of as separating regions of conducting and insulating states. However, our result shows that in AAH model with power-law hopping one can have a phase, where there is a mobility edge separating two different kinds of conducting states, viz., ballistic and super-diffusive. To our knowledge, this is the first time the possibility of such a system is being reported.

We have demonstrated this here taking the irrational number $b$ as the golden mean. But the same physics holds for other choices of irrational numbers. Any irrational number can be expanded in an infinite continued fraction. Truncating the continued fraction at any stage gives a rational approximation to the irrational number. Truncating at various levels of the continued fraction, a series of rational approximations to the irrational number can be obtained. The near resonance condition will then occur for sites separated by a distance equal to the denominators of the rational approximations. For the golden mean, these numbers are the Fibonacci numbers.

While the physics described above is immune to the choice of the irrational number, it is completely due to the quasiperiodic nature of the potential. So, instead of the AAH potential, if the system had random disorder there would not be the secondary peaks. In fact, it is known that in such cases, for algebraically localized states, $|\Phi_{n}(x^\prime)|^2_{typ}\sim {x^{\prime}}^{-2\alpha}$ \cite{theory_power_law_loc1,theory_power_law_loc3}. Thus, for random disorder, from Eq.~\ref{alg_loc_criterion}, the algebraically localized states will be insulating for $\alpha>1$. As a result, localization due to random disorder and localization due to quasiperiodic disorder leads to extremely different transport properties in presence of power-law hopping.

\subsection{Summary and outlook}
Let us now summarize all the main results in this paper. In Sec.~\ref{theory}, we have analytically explored on general grounds the relation between localization and nature of transport for algebraically localized states. We have proven that a non-interacting fermionic system at zero temperature is insulating if and only if the mean position of a particle at Fermi energy is well-defined in the thermodynamic limit.  Based on this criterion, when Fermi level corresponds to an algebraically localized state, the system may be either conducting or insulating depending on the strength of the algebraic decay. In Sec.~\ref{numerical_example}, we have given such an example.

In Sec.~\ref{numerical_example}, the numerical example we have considered is a system with AAH potential in presence of power-law hopping. We have chosen the parameters of the AAH potential such that in absence of power-law hopping, the single-particle eigenstates would be exponentially localized ($W>2$). In presence of power-law hopping, as shown in \cite{theory_power_law_loc6}, there is a mobility edge separating completely delocalized states and algebraically localized states. We have shown that, due to the quasiperiodic nature of the AAH potential, there occurs near-resonance conditions, which causes a series of secondary peaks in typical algebraically localized eigenfunctions. The algebraic decay of the height of these peaks is such that, for $1<\alpha\leq2$, the mean of the probability distribution given by the square of the eigenfunction is ill-defined in the thermodynamic limit. For $\alpha>2$, the corresponding mean is well-defined in the thermodynamic limit. Thus, at zero temperature, when the Fermi level corresponds to an algebraically localized state, the system is conducting for $1<\alpha\leq2$, while it is insulating for $\alpha>2$. Classifying transport in terms of the zero temperature Drude weight and the zero temperature many-particle localization length, we have shown that the algebraically localized states, for $1<\alpha<2$, lead to super-diffusive transport.  Thus, for $W>2$ and $1<\alpha<2$, we have found a phase where there is a mobility edge which separates two different kinds of conducting states, ballistic and super-diffusive. This is in contrast with general wisdom, where mobility edges are usually thought of as separating conducting and insulating states.

Our work opens several new questions regarding quasi-periodic one-dimensional systems with power-law hopping and points to the rich physics of such systems. For the AAH model with power-law hopping, in this work, we have only explored a part of the phase diagram in terms of transport properties. It has been previously shown that even when the short-ranged AAH model is delocalized ($W<2$), switching on power-law hopping can lead to localization \cite{theory_power_law_loc4}. This falls in the paradigm of the recently discussed `correlation induced localization' \cite{theory_power_law_loc3}. It is of interest to explore the transport through such localized states in the light of our results. Definitely, the mechanism for localization or delocalization will be different for such states. The case of the critical AAH model ($W=2$) in presence of long-range hopping, deserves to be studied even more thoroughly and there has been almost no work exploring this. Further, it has recently been shown that isolated system transport properties and open system transport properties can be extremely different for quasi-periodic systems \cite{Archak_phase_diagram,Archak_AAH,Archak_AAH1}. Thus, the open system transport properties of quasi-periodic one-dimensional systems with power-law hopping is also of extreme interest. In this work, we have shown the existence of a  single-particle mobility edge that separates regions of ballistic and super-diffusive transport. The effect of interactions on such a mobility edge is also one of the interesting directions to explore, which may be experimentally possible in trapped ion experiments, a platform where long-ranged quasi-periodic systems may be possible to engineer\cite{Experiment_transport}. From a practical point of view, such systems, with extremely rich and tunable transport properties, may find use in devising rectifiers \cite{rectification1,rectification2,rectification3} and autonomous quantum heat-engines \cite{thermal_machines_review}.

\section*{ACKNOWLEDGEMENT}
MS would like to acknowledge University Grants Commission (UGC) of India for her research fellowship. SKM would like to acknowledge the financial support of DST-SERB, Government of India (Project File Number: EMR/2017/000504). AP acknowledges funding from the European Research Council (ERC) under the European Unions Horizon 2020 research and innovation program (grant agreement No. 758403).

\section*{Appendix}
\appendix

\section{Kubo conductivity under open boundary conditions}\label{Kubo_deriv}
Here, we outline the steps for derivation of Eq.~\ref{Kubo_obc}.
The Kubo formula for particle conductivity of an isolated system in the thermodynamic limit has the form 
\begin{align}
\label{Kubo}
\sigma (\omega) = \pi D(N) \delta (\omega)+ \sigma^{reg}(\omega).
\end{align}
Let us consider a many-body fermionic system in a 1D lattice of $N$ sites with Hamiltonian given by
\begin{align}
\hat{\mathcal{H}}=\sum_{n} E_n |n\rangle\langle n|,
\end{align}
where $|n\rangle$ is a many-body eigenstate of the Hamiltonian with energy $E_n$. In this case, the expressions for $D$ and $\sigma^{reg}(\omega)$ are,
\begin{align}
D &= \lim_{N\rightarrow\infty} D(N) \nonumber \\
D(N) &=\frac{i}{N}\Big(\langle~[\hat{I},\sum x \hat{n}_x ]~\rangle \nonumber \\
&-\sum_{E_n\neq E_m} \frac{p_n-p_m}{E_m-E_n} |\langle n|\hat{I}|m\rangle|^2\Big) \nonumber \\
\sigma^{reg}(\omega)&= \lim_{N\rightarrow\infty} \frac{\pi}{N} \frac{(1-e^{-\beta \omega})}{\omega} \sum_{E_n\neq E_m} \Big[ \nonumber \\
& p_n |\langle n|\hat{I}|m\rangle|^2 \delta (\omega-E_m+E_n)\Big],
\end{align}
where $p_n=\exp(-\beta E_n)/Z$, $Z=\sum_n \exp(-\beta E_n)$, $\langle ... \rangle=Tr(\exp(-\beta \hat{\mathcal{H}})/Z ...)$, $\hat{n}_x=\hat{c}_x^\dagger \hat{c}_x$ is the local particle density operator and $\hat{I}$ is the particle current operator.

Under open boundary condition, the particle current operator is given by 
\begin{align}
\label{def_x_I}
& \hat{I}=\frac{d\hat{x}}{dt}=-i [\hat{x},\hat{\mathcal{H}}], \nonumber \\
& \textrm{where } \hat{x}=\sum_{x} x \hat{n}_x
\end{align}
is the position operator. With this definition of particle current operator, it can be checked that $D(N)=0$. As shown in \cite{Rigol_drude}, if transport is ballistic in such case, then $\sigma^{reg}(\omega)$ develops a peak at finite frequency, which grows in height and moves towards zero frequency as system-size is increased. In this way, Eq.~\ref{Kubo} is recovered in the thermodynamic limit. So under open boundary condition, we have
\begin{align}
&\sigma (\omega) = \lim_{N\rightarrow\infty} \frac{\pi}{N} \frac{(1-e^{-\beta \omega})}{\omega} \sum_{E_n\neq E_m}  \Big[ \nonumber \\
& p_n \Big| \sum_{q} q (E_m-E_n)\langle n|\hat{n}_q|m\rangle\Big|^2 \delta (\omega-E_m+E_n)\Big].
\end{align}  
Converting the $\delta$-function to an integral and noting that only $\omega=E_m-E_n$ contributes, we can rewrite above expression as
\begin{align}
&\sigma (\omega) =  \int_{-\infty}^{\infty} dt\lim_{N\rightarrow\infty}  \frac{\pi}{N} \sum_{n,m}\Big[ e^{i(\omega-E_m+E_n)t} \nonumber \\
&  (p_n-p_m)(E_m-E_n)\Big| \sum_{q} q \langle n|\hat{n}_q|m\rangle\Big|^2 \Big].
\end{align}
That above equation is same as Eq.~\ref{Kubo_obc} can be checked by directly evaluating Eq.~\ref{Kubo_obc} in the many-particle basis.

\section{Classification of transport}\label{classification}
Here we give details of the standard way to classify transport into ballistic, super-diffusive, diffusive, sub-diffusive and exponentially localized (absence of diffusion). To do this, we re-write Eq.~\ref{Kubo_obc}
\begin{align}
\label{Kubo_obc2}
\sigma(\omega)=&\pi\int_{-\infty}^{\infty} dt e^{i\omega t} \nonumber\\
&\frac{d}{dt}\left( \lim_{N\rightarrow\infty} \frac{i}{N}\sum_{p,q=-\ceil{N/2}}^{\ceil{N/2}} pq \langle [\hat{n}_p(t), \hat{n}_q(0)] \rangle\right).
\end{align}
This equation shows that long time behavior of the following quantity governs low-frequency behavior of $\sigma(\omega)$,
\begin{align}
&\frac{d}{dt}\left( \lim_{N\rightarrow\infty} \frac{i}{N}\sum_{p,q=-\ceil{N/2}}^{\ceil{N/2}} pq \langle [\hat{n}_p(t), \hat{n}_q(0)] \rangle\right)\nonumber \\
&=\frac{d}{dt}\left( \lim_{N\rightarrow\infty} \frac{-i}{2N}\sum_{p,q=-\ceil{N/2}}^{\ceil{N/2}} (p-q)^2 \langle [\hat{n}_p(t), \hat{n}_q(0)] \rangle\right) \nonumber \\
\end{align}
In going from the first line to the second, we have used $2pq=p^2+q^2-(p-q)^2$, and have noted that $\sum_p \hat{n}_p=N_e$ is the total number of particles, which is a conserved quantity.  So we see that the long time behavior of $\langle [\hat{n}_p(t), \hat{n}_q(0)] \rangle$ governs the nature of transport. Using standard linear response theory, $\langle [\hat{n}_p(t), \hat{n}_q(0)] \rangle$ can be interpreted as being proportional to the linear response of the system at site $p$ and time $t$, when a small instantaneous perturbation is given at site $q$ at time $t=0$. So $\langle [\hat{n}_p(t). \hat{n}_q(0)] \rangle$ quantifies `diffusion' of an initial instantaneous perturbation. Let the long time behavior be such that, for $t\gg 1$,
\begin{align}
\label{transport_and_diffusion}
\lim_{N\rightarrow\infty}\left(\frac{1}{2N}\sum_{p,q=-\ceil{N/2}}^{\ceil{N/2}} (p-q)^2 \langle [\hat{n}_p(t), \hat{n}_q(0)] \rangle \right)\sim t^{\tilde{s}}.
\end{align}
Then from Eq.~\ref{Kubo_obc2}, we see that, by property of Fourier transform, the low-frequency behavior $\sigma(\omega)$ is given by
\begin{align}
\label{s_and_tilde_s}
\sigma(\omega)\sim \omega^{-s},~~s=\tilde{s}-1
\end{align}
For diffusive spread of initial perturbation $\tilde{s}\sim 1$, for ballistic spread $\tilde{s}\sim 2$, for absence of diffusion $\tilde{s}\sim \leq 0$, in which the RHS of Eq.~\ref{transport_and_diffusion} goes to a constant. This leads to the following classification of transport,
\begin{align}
&\tilde{s}=2,~~\textrm{ballistic transport,} \nonumber \\
&1<\tilde{s}<2,~~\textrm{super-diffusive transport,} \nonumber \\
&\tilde{s}=1, ~~\textrm{diffusive transport,} \\
&0<\tilde{s}<1, ~~\textrm{sub-diffusive transport,} \nonumber \\
&\tilde{s}\leq 0, ~~\textrm{absence of diffusion, exponentially localized.} \nonumber
\end{align}
The corresponding values of $s$ can be found from Eq.~\ref{s_and_tilde_s}. It is clear from above that for in ballistic and super-diffusive cases, the system is conducting, with a diverging conductivity, in the diffusive case, it has a finite conductivity, in sub-diffusive and exponentially localized cases, the system is insulating. Note that in the sub-diffusive case there is diffusion but the system is insulating.

\section{Many particle localization length from $\sigma^{reg}$}
\label{xi_Kubo_reg}
Here we give the derivation of Eq.~\ref{def_xi}.
Without loss of generality, we will assume that the many-body ground state energy $E_0=0$. At zero temperature, then, we have,
\begin{widetext}
\begin{align}
 \lefteqn{ \lim_{N\rightarrow\infty} \int_{\omega_{min}}^{\omega_{max}} \frac{\sigma^{reg}(\omega)}{\omega} d\omega} \\ \nonumber
&= \lim_{N\rightarrow\infty} \frac{\pi}{N} \int_{0}^{\infty} \frac{(1-e^{-\beta \omega})}{\omega^2} \sum_{E_n\neq E_m} p_n |\langle n|\hat{I}(0)|m\rangle|^2 \delta (\omega-(E_m-E_n)) d\omega   \\ \nonumber
&=\lim_{N\rightarrow\infty} \frac{\pi}{N}\sum_{E_m \neq E_n} \frac{p_n (1-e^{-\beta (E_m-E_n)})}{(E_m-E_n)^2} |\langle n|\hat{I}(0)|m\rangle|^2 \\ \nonumber
&\propto \lim_{N\rightarrow\infty} \frac{1}{N}\sum_{E_m \neq E_0}  |\langle 0|\hat{x}(0)|m\rangle|^2 ~~~~(put ~\beta\rightarrow\infty~and~\langle n|\hat{I}(0)|m\rangle|=-i(E_m-E_n)\langle n|\hat{x}(0)|m\rangle|)\\ \nonumber
&~~(P_n=1, ~as~ the~ contribution ~will~ come~ from~ the~ ground state ~(n=0)~only~ at~zero~temperature.) \\ \nonumber
& In~ previous~ line,~ the ~sum~ is~ over~ all~ m~ except~ the~ ground state.~ This~ can ~be~ written~ as~\\ \nonumber
& deducting ~the~ term~(E_m=E_0)~ from ~sum ~over~ all ~the~ values ~of~ m.~ This ~then~becomes~,\\ \nonumber
&\propto \lim_{N\rightarrow\infty}\frac{1}{N} \langle 0|\hat{x}^2(0)|0\rangle-\Big|\langle 0|\hat{x}(0)|0\rangle\Big|^2\Big) \\ \nonumber
&\propto \lim_{N\rightarrow\infty}\frac{1}{N} \Big(\langle \hat{x}^2(0)\rangle-\langle \hat{x}(0)\rangle^2\Big)\propto \lim_{N\rightarrow\infty} \xi^2(N)
\end{align}
\end{widetext}
Thus we have obtained Eq.~\ref{def_xi}.

\section{Condition for existence of RHS of Eq.~\ref{Kubo_and_mean}}\label{RHS_existence}
The existence of RHS of Eq.~\ref{Kubo_and_mean} requires the following limit, 
\begin{align}
\label{RHS}
\lim_{N\rightarrow \infty} \frac{\overline{x}(E_F)}{\sqrt{N}}
\end{align}
to be well-defined. It is obvious that sufficient condition for is that $\lim_{N\rightarrow\infty}\overline{x}(E_F)$ is well-defined. This requires that the two following limits exist independently (see Eq.~\ref{mean_existence}),
\begin{align}
\label{mean_existence_condition}
\lim_{b\rightarrow\infty} \sum_{-a}^{b} x P_n(x), ~\forall~\textrm{finite a} \nonumber \\
\lim_{a\rightarrow\infty} \sum_{-a}^{b} x P_n(x), ~\forall~\textrm{finite b}.
\end{align}
We will show here that the above is also a necessary condition for Eq.~\ref{RHS} to  be well-defined. To see this, observe that Eq.~\ref{RHS} is well-defined only if
\begin{align}
\lim_{N\rightarrow \infty} \frac{\overline{x}(E_F)}{\sqrt{N}} &= \lim_{a\rightarrow\infty}\frac{1}{\sqrt{a}}\left[\lim_{b\rightarrow\infty} \sum_{-a}^{b} x P_n(x)\right] \nonumber \\
&= \lim_{b\rightarrow\infty}\frac{1}{\sqrt{b}}\left[\lim_{a\rightarrow\infty} \sum_{-a}^{b} x P_n(x)\right].
\end{align}
All limits in the above equation must exist. This clearly shows that the limits in Eq.~\ref{mean_existence_condition} must exist for Eq.~\ref{RHS} to be well-defined. So, the existence of $\lim_{N\rightarrow\infty}\overline{x}(E_F)$ is a necessary and sufficient condition for Eq.~\ref{RHS} to be well-defined. If Eq.~\ref{RHS} is well-defined, then its value is 0.

\section{Existence of thermodynamic limit}\label{appendix:therm_limit}
\begin{figure}[!h]
\includegraphics[width=\columnwidth]{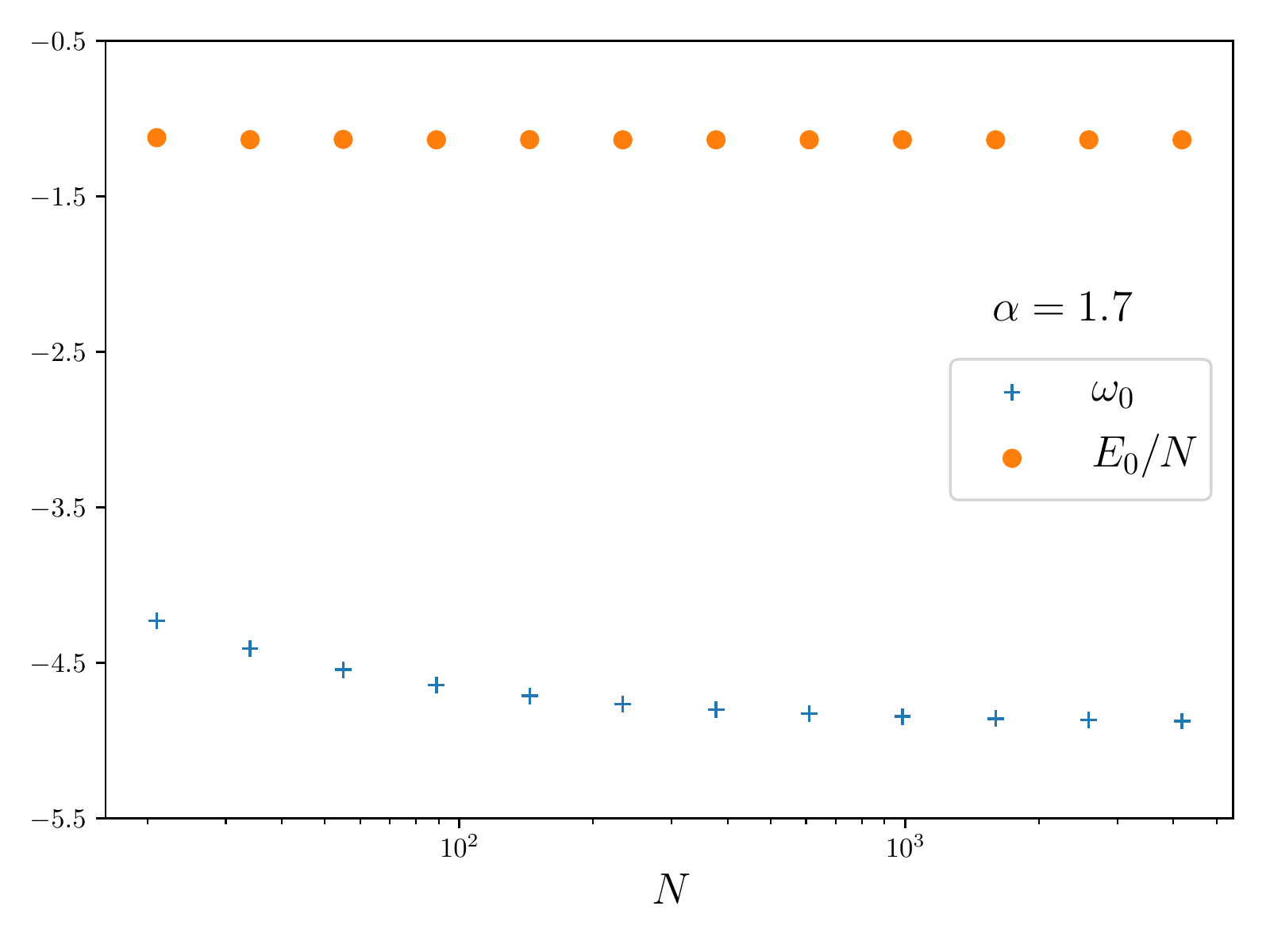}
\caption{(Color online) Variation of the energy the lowest single-particle level $\omega_0$ and the ground state energy $E_0$ at half-filling with $N$ for $\alpha=1.7$. Both $\omega_0$ and $E_0/N$ reach to a constant with increase in $N$. $W=3$.} 
\label{fig:therm_limit_check}
\end{figure}

We have explicitly checked the existence of thermodynamic limit for our model Hamiltonian Eq.~\ref{AAHlr} for $\alpha>1$. For this, we look at the variation of the energy of the lowest single-particle level $\omega_0$, and the ground state energy $E_0$ with system-size $N$ at a fixed filling. If $\omega_0$ and $E_0/N$ both reach to constant, then the thermodynamic limit is well-defined. In Fig.~\ref{fig:therm_limit_check}, we show plots of $\omega_0$ and $E_0/N$ at half-filling for $\alpha=1.7$. It is clear that the thermodynamic limit exists. Though we present a plot here for half-filling, we have checked that this remains true for any fixed filling, and for all $\alpha>1$.

\section{Algebraically and exponentially localized eigenfunctions}\label{appendix:alg_exp_loc}
\begin{figure}[!h]
\includegraphics[width=\columnwidth]{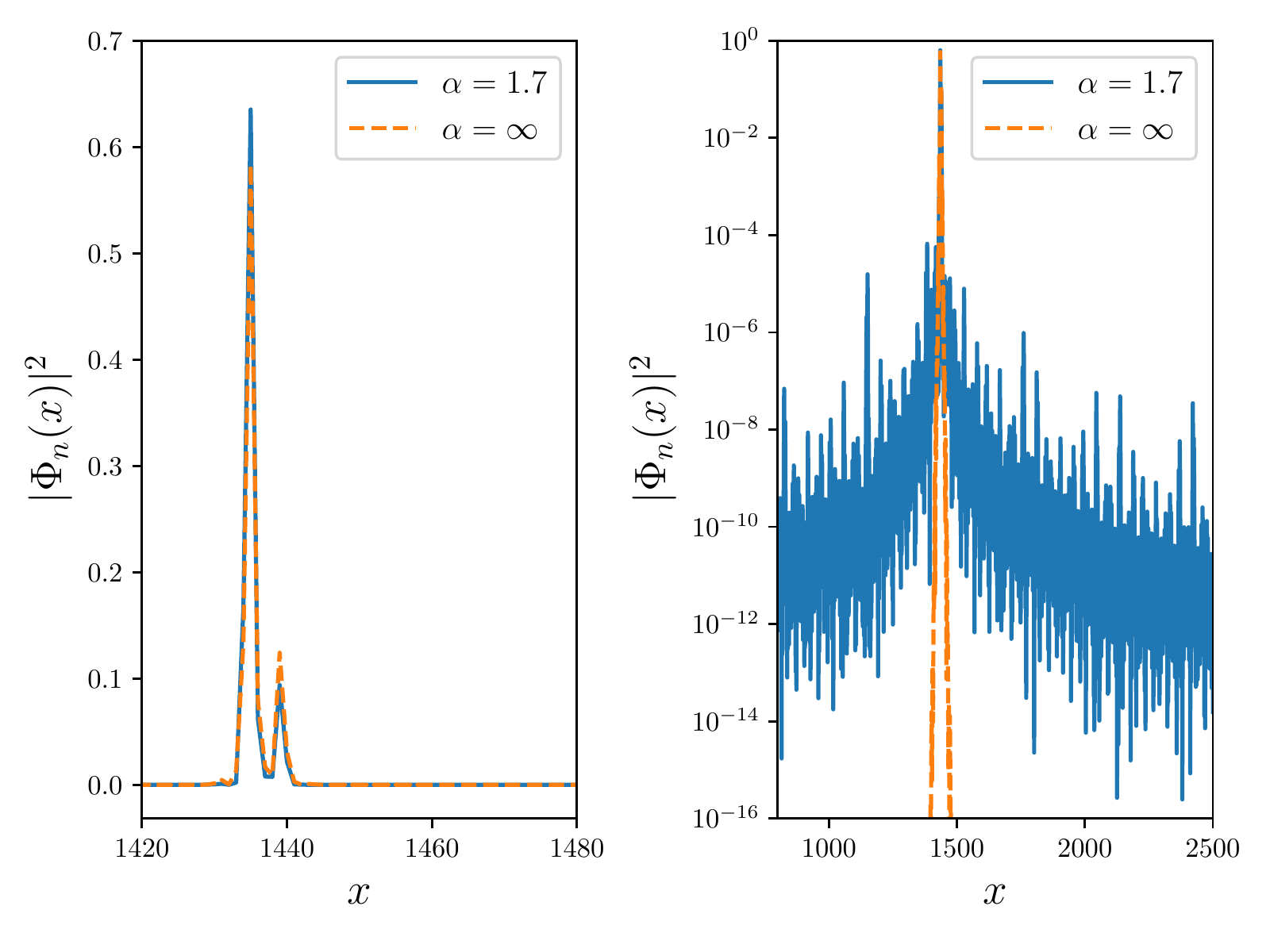}
\caption{(Color online) The figure shows plots of $|\Phi_{n}(x)|^2$ for $n=2000$ for the power-law decaying system ($\alpha=1.7$), and the nearest neighbour hopping system ($\alpha=\infty$) with $W=3$. The left panel shows the plot in linear scale, while the right panel shows the same plot with y-axis in log-scale. $N=4181$.} 
\label{fig:alg_and_exp_loc_wvfcns}
\end{figure}
\begin{figure*}
\includegraphics[width=\textwidth, height=7cm]{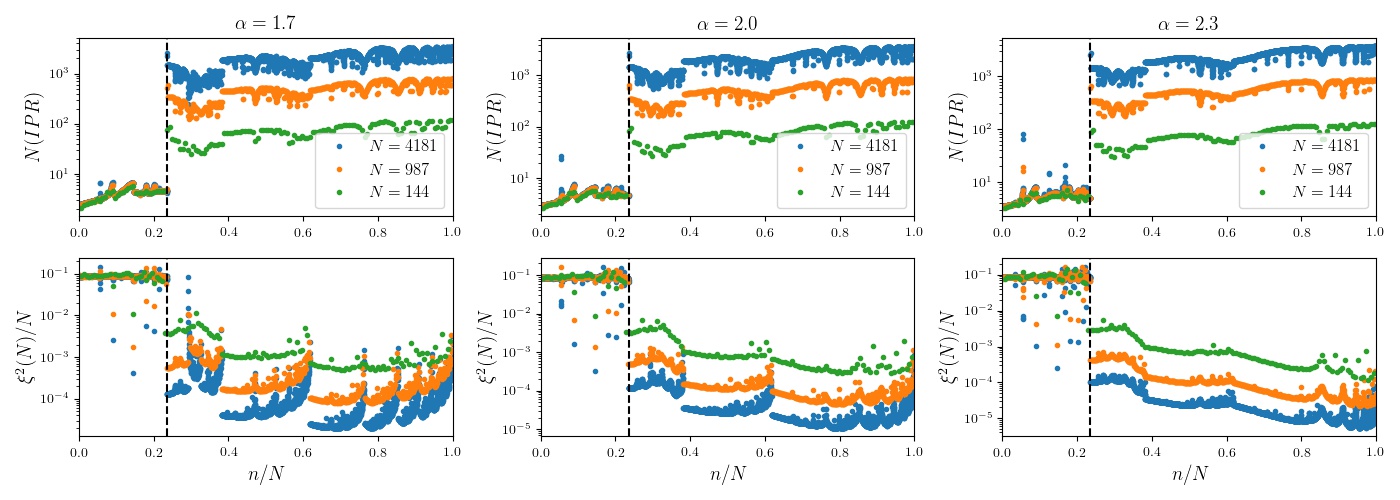}
\caption{(Color online)  This figure is complementary to Fig.~\ref{fig:IPR_Drude_mpll_3_values} with $IPR$ multiplied by $N$ and $\xi^2$ divided by $N$ so show the scaling corresponding to the completely delocalized states for these quantities.  $W=3$.} 
\label{fig:IPR_mpll_deloc_scaled}
\end{figure*}
Here we explicitly show the difference between algebraically localized and exponentially localized eigenfunctions by plotting them on the same axis. For this purpose, we compare the localized states of the nearest neighbour AAH model ($\alpha=\infty$) with those of the AAH model with power-law decay. Figure.~\ref{fig:alg_and_exp_loc_wvfcns} shows plots of the $2000$th single-particle eigenfunction for $W=3$ and $\alpha=1.7$ and for the nearest neighbour AAH model. The left panel of the Fig.~\ref{fig:alg_and_exp_loc_wvfcns} shows the eigenstates in the linear scale. The eigenstates of the two different models seem to overlap. Thus, their peaks are at the same position, and the height of the peaks are nearly the same. This leads to having almost same $IPR$ values. The right panel of the Fig.~\ref{fig:alg_and_exp_loc_wvfcns} shows the same plots with y-axis in log-scale. It is completely clear that the eigenfunction corresponding to $\alpha=1.7$ decays algebraically, while that of the nearest neighbour model decays exponentially. Thus their tails are very different. As we have shown in the main-text, this leads to very different transport behavior, for $\alpha=1.7$, the `localized' states are conducting, while for the nearest neighbour model, they are known to be insulating.

\section{Scaling for the delocalized states}\label{appendix:scaling_deloc}

In Fig.~\ref{fig:IPR_Drude_mpll_3_values} of the main text we have given the plots of $IPR$, $D(N)$ and $\xi^2$ as a function of $n/N$. We have mentioned that scaling of the $IPR$ for the completely delocalized states can be confirmed by multiplying the data points by $N$, while that for $\xi^2$ can be confirmed by dividing the data points by $N$. Here, in Fig.~\ref{fig:IPR_mpll_deloc_scaled}, we show this by plotting $N~(IPR)$ and $\xi^2/N$ with $n/N$ for the chosen values of $\alpha$. It is clear that, for $n/N<b^3$, the data points for $N~(IPR)$ and $\xi^2/N$  for various system sizes collapse. Thus, for this case, $IPR \sim 1/N$ and $\xi^2 \sim N$, as expected for completely delocalized states and ballistic transport. One can observe slight deviations at few points, especially for $\xi^2/N$. These are due to finite size effects, and goes away as system-size is increased.

\section{The average algebraically localized eigenfunction}\label{appendix_avg_phi}
\begin{figure}[!h]
\includegraphics[width=\columnwidth]{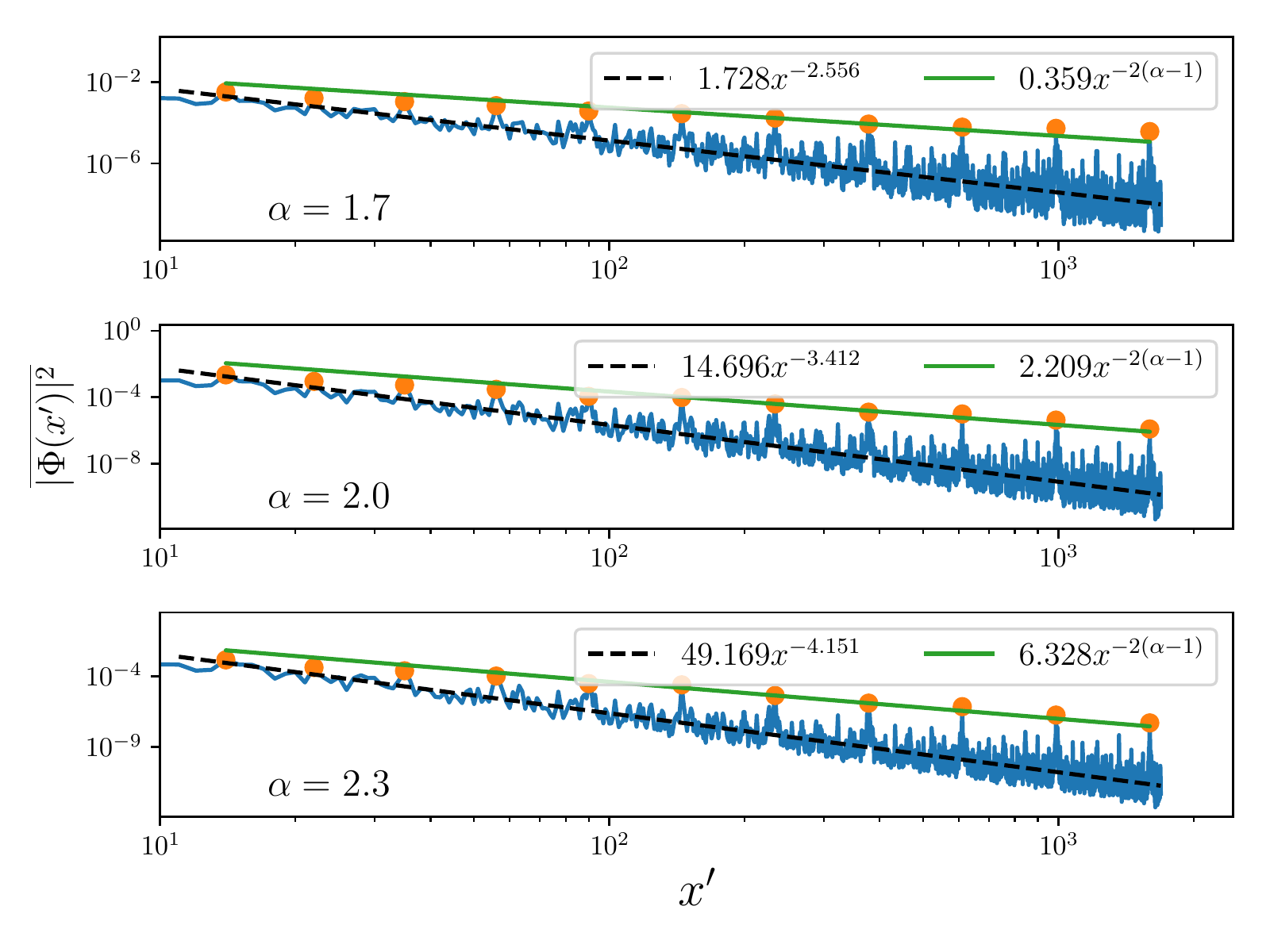}
\caption{(Color online) The figure shows scaling of $\overline{|\Phi_{n}(x^\prime)|^2}$ with $x^\prime$ for $\alpha=1.7$ (top), $\alpha=2.0$ (middle), $\alpha=2.3$ (bottom). The orange dots correspond to points where $x^\prime=$Fibonacci number. The green line shows the scaling of  $|\Phi_{n}(x^\prime)|^2_{typ}$ for these points. The black dashed line shows a least square fit of all the data points. System size: $N=6765$.  $W=3$.} 
\label{fig:tail_scaling_avg}
\end{figure}
In the main text, we have looked at the scaling of tails of the typical algebraically localized eigenfunction given by geometric mean of all the algebraically localized eigenfunctions (Eq.~\ref{def_phi_typ}). Here we look at the arithmetic mean of all algebraically localized eigenfunctions,
\begin{align}
&\overline{|\Phi_{n}(x^\prime)|^2} = \frac{1}{\mathcal{N}}\sum_{n}^\prime |\Phi_{n}(x^\prime)|^2, \\
& x^\prime = x-x_0,~~x>x_0, \nonumber
\end{align} 
where $x_0$ is the position of the peak of the algebraically localized state, $\mathcal{N}$ is the number of algebraically localized states, $\sum_{n}^\prime$ denotes sum over all algebraically localized states. Compared to the $|\Phi_{n}(x^\prime)|^2_{typ}$, $\overline{|\Phi_{n}(x^\prime)|^2}$ is expected to show more finite-size effects. This is because, atypical behavior due to finite-size can make a considerable contribution to 
$\overline{|\Phi_{n}(x^\prime)|^2}$, while those are suppressed in $|\Phi_{n}(x^\prime)|^2_{typ}$. Figure.~\ref{fig:tail_scaling_avg} shows plots of 
$\overline{|\Phi_{n}(x^\prime)|^2}$ for the exact same parameters as for $|\Phi_{n}(x^\prime)|^2_{typ}$ in Fig.~\ref{fig:tail_scaling}. Due to finite-size effects, the least square fit of all data points do not seem to decay with an exponent $\sim 2\alpha$, which was seen for $|\Phi_{n}(x^\prime)|^2_{typ}$. Nevertheless, the peaks for $x^\prime=F_n$ still exist, with
\begin{align}
\overline{|\Phi_{n}(x^\prime)|^2} &\sim F_{n}^{-2(\alpha-1)},
\end{align} 
though the scaling is slightly worse than for $|\Phi_{n}(x^\prime)|^2_{typ}$. The scaling seems to become better at larger system sizes, as expected. Thus, both the geometric mean and the arithmetic mean give the same conclusion. This conclusively shows that the secondary peaks at $x^\prime=F_n$  is indeed the generic behavior of the algebraic localized eigenfunctions of the AAH model with power-law hopping, and not any artefact of any averaging procedure.

\bibliography{ref_long_range}

\end{document}